\documentclass[lettersize, journal]{IEEEtran}

\usepackage{amsmath, amssymb, amsfonts, amssymb, amsthm}

\usepackage{bbm}
\usepackage{graphicx} \graphicspath{ {figure/} }
\usepackage{epstopdf} 
\usepackage[noabbrev]{cleveref}
\usepackage{scalefnt}
\usepackage{indentfirst}
\usepackage{cite}
\usepackage{bm}
\usepackage{enumerate}
\usepackage{enumitem}
\usepackage[caption=false,subrefformat=parens,labelformat=parens]{subfig}
\usepackage{scalerel}
\usepackage{scalerel}
\usepackage{algpseudocode}
\usepackage{etoolbox}
\usepackage{multirow}
\usepackage[table,xcdraw]{xcolor}

\usepackage{algorithm}
\usepackage{algcompatible}
\renewcommand{\algorithmiccomment}[1]{\hfill ~#1}
\algnewcommand\INPUT{\item[\textbf{Input:}]}
\algnewcommand\INITIAL{\item[\textbf{Initialization:}]}
\algnewcommand\OUTPUT{\item[\textbf{Output:}]}
\algnewcommand\RETURN{\item[\textbf{Return:}]}
\algnewcommand\ITER{\item[\textbf{Iteration:}]}
\algrenewcommand\algorithmiccomment[2][\small]{{#1\hfill\ #2}}

\theoremstyle{plain}

\makeatletter
\newcommand*{\algrule}[1][\algorithmicindent]{%
  \makebox[#1][l]{%
    \hspace*{.2em}
    \vrule height .75\baselineskip depth .25\baselineskip
  }
}

\newcount\ALG@printindent@tempcnta
\def\ALG@printindent{%
    \ifnum \theALG@nested>0
    \ifx\ALG@text\ALG@x@notext
    \else
    \unskip
    \ALG@printindent@tempcnta=1
    \loop
    \algrule[\csname ALG@ind@\the\ALG@printindent@tempcnta\endcsname]%
    \advance \ALG@printindent@tempcnta 1
    \ifnum \ALG@printindent@tempcnta<\numexpr\theALG@nested+1\relax
    \repeat
    \fi
    \fi
}
\patchcmd{\ALG@doentity}{\noindent\hskip\ALG@tlm}{\ALG@printindent}{}{\errmessage{failed to patch}}
\patchcmd{\ALG@doentity}{\item[]\nointerlineskip}{}{}{} 
\makeatother

\renewcommand{\algorithmiccomment}[2][.5\linewidth]{\leavevmode\hfill\makebox[#1][l]{//~#2}}

\theoremstyle{definition}

\begin{document}
\title{Large Multimodal Model-aided Scheduling for 6G Autonomous Communications}

\author{Sunwoo Kim, \textit{Member}, \textit{IEEE}, and Byonghyo Shim, \textit{Fellow}, \textit{IEEE}\\
\IEEEauthorblockA{Institute of New Media and Communications, Department of Electrical and Computer Engineering, Seoul National University, Seoul, Korea\\
Email: swkim@islab.snu.ac.kr, bshim@snu.ac.kr}

\thanks{A part of this paper has been presented at IEEE Vehicular Technology Conference (VTC), Chengdu, China, Oct. 2025~\cite{conference}.
This work was supported in part by Institute of Information \& communications Technology Planning \& Evaluation (IITP) under the Graduate School of Artificial Intelligence Semiconductor(IITP-2025-RS-2023-00256081) grant funded by the Korea government(MSIT), in part by the National Research Foundation of Korea(NRF) grant funded by the Korea government(MSIT) (2022M3C1A3099336), and in part by the MSIT(Ministry of Science and ICT), Korea, under the ITRC(Information Technology Research Center) support program(IITP-2023-2021-0-02048) supervised by the IITP(Institute for Information \& Communications Technology Planning \& Evaluation).
}

}

\maketitle

\begin{abstract}
Recently, large language models (LLMs) have gained significant attention for their ability to generate fast and accurate answer to the given query. These models have evolved into large multimodal models (LMMs), which can interpret and analyze multimodal inputs such as images and text.
With the exponential growth of AI functionalities in autonomous devices,  the central unit (CU), a digital processing unit performing AI inference, needs to handle LMMs to effectively control these devices. To ensure seamless command delivery to devices, the CU must perform the scheduling, which involves resource block (RB) allocation for data transmission and modulation and coding scheme (MCS) index selection based on the channel conditions. 
This task is challenging in many practical environments in 6G, where even small user movement can cause abrupt channel changes. In this paper, we propose a novel LMM-based scheduling technique to address this challenge. 
Our key idea is to leverage LMM to predict future channel parameters (e.g., distance, angles, and path gain) by analyzing the visual sensing information as well as pilot signals. 
By exploiting LMMs to predict the presence of reliable path and geometric information of users from the visual sensing information, and then combining these with past channel states from pilot signals, we can accurately predict future channel parameters.
Using these predictions, we can preemptively make channel-aware scheduling decisions.
From the numerical evaluations, we show that the proposed technique achieves more than 30\% throughput gain over the conventional scheduling techniques.
\end{abstract}

%

\section{Introduction}\label{sec:intro}
Recently, generative AI has received great attention for its capability to create high-quality outputs that even humans cannot easily generate~\cite{generativeAI}.
Among various forms of generative AI, large language models (LLMs) have garnered significant attention for their capacity to answer questions without requiring additional examples or hints (i.e., zero-shot prompting) and solve problems with only a few input-output pairs (i.e., few-shot prompting)~\cite{LMM}. 
While LLMs have been primarily focused on language data, they have recently evolved into large multimodal models (LMMs) to handle and interpret diverse modalities such as images, videos, and audio.
Due to their flexibility and adaptability, LMMs are now actively integrated into automation applications such as driverless vehicles~\cite{autonomousdrivingLMM}, humanoids~\cite{humanoidLMM}, and surgical/agricultural manufacturing~\cite{surgeryLMM}.

\begin{figure}[t]
	\centering
\includegraphics[width=1.0\columnwidth, height=5.0cm]{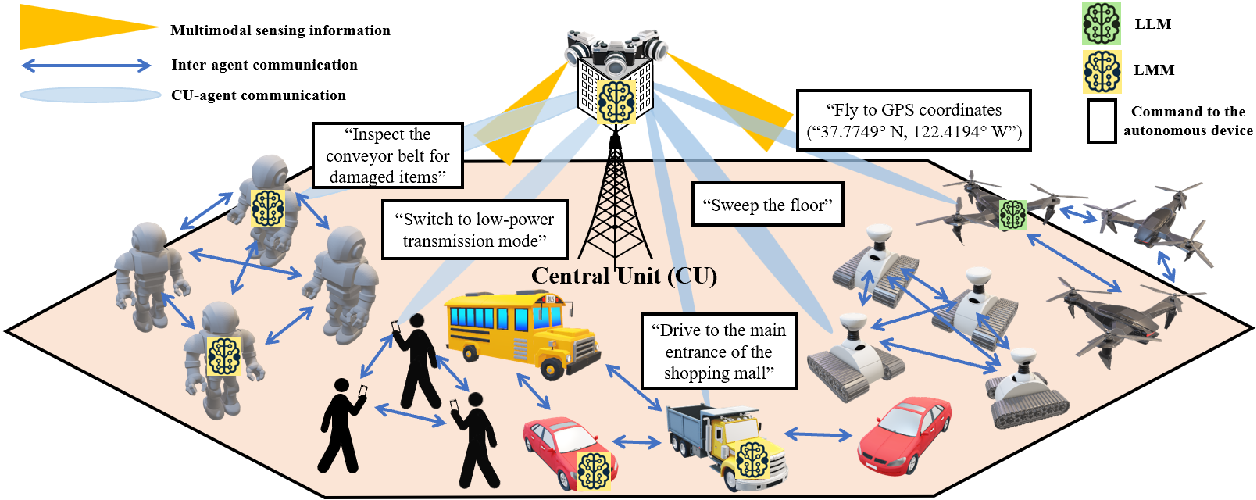}
         \vspace{-0.5cm}
        \caption{LLM/LMM-based task-oriented communication system.}
	\label{fig:LLM}
    \vspace{-0.7cm}
\end{figure}
While there are lots of trials to integrate LLMs/LMMs into  autonomous tasks (e.g., robot, humanoid, and autonomous vehicle), their application to future
 wireless communication systems is still relatively scarce~\cite{motivation1}. This is mainly because AI operations
 are often confined in a single agent (e.g., a humanoid) so that the exchange of information with other agents and the central unit (CU) (e.g., cloud center or digital unit) is unnecessary, or at least not that important.
 In the future, however, as the range of tasks performed
 by autonomous devices grows exponentially, far more agents will cooperate and collaborate, making both inter-agent communication and communication between the CU and agents critical~\cite{motivation2}.
 In such systems, wireless
 communications will play a central role for exchanging status information among agents to provide mutual feedback, as well as multimodal sensing information (e.g., LiDAR and radio frequency (RF) signal) to control these agents using the LMM in the CU (see Fig. 1).



\begin{figure*}[t]
	\centering
    \includegraphics[width=2.0\columnwidth, height=6.0cm]{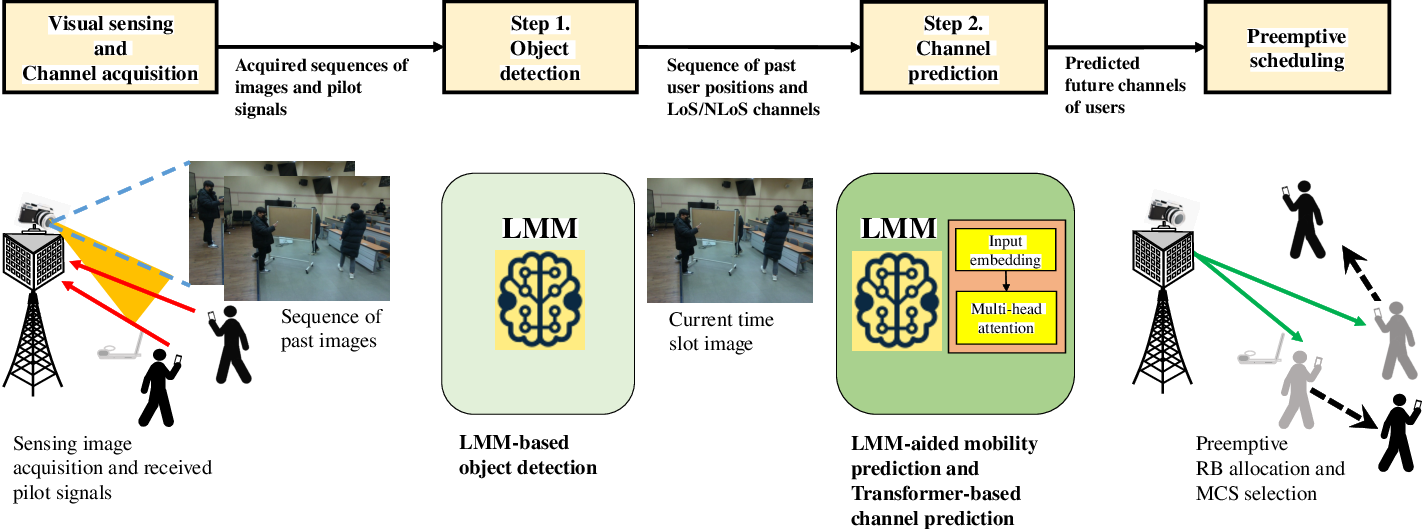}
        \vspace{-0.5cm}
        \caption{Conceptual illustration of the proposed LMM-PS.}
	\label{fig:overall}
    \vspace{-0.5cm}
\end{figure*}
In order to ensure that autonomous devices can perform their designated operations such as car assembly or remote surgery, CU needs to perform the scheduling, a task to allocate resource blocks (RBs) for the data transmission and select appropriate link parameters (i.e., modulation and coding scheme (MCS)) based on the channel conditions. 
For instance, by selecting a proper MCS index conforming to the channel quality, CU can satisfy the symbol error rate (SER) requirement while ensuring seamless command delivery to  devices.
In the upcoming 6G environments, however, scheduling will be by no means easy due to sudden channel state changes\footnote{Even slight user movements can cause the line-of-sight (LoS) path to disappear, leading to a significant signal-to-noise ratio (SNR) drop (e.g., up to 35\,dB in the mmWave band~\cite{YJAhn_SNRdrop}) when shifting from LoS to non-LoS (NLoS).} caused by densely distributed obstacles, various spectrums, and also movement of autonomous devices.
In such scenarios, conventional scheduling techniques relying on outdated channel information might fail to meet users' quality-of-service (QoS) requirements, resulting in an operation failure.

To overcome the problem, one can consider a preemptive scheduling strategy that performs the scheduling decision based on predicted future channel conditions~\cite{conv1_preemptive, conv2_preemptive, conv3_preemptive}.
The core idea of this strategy is to capture the temporal relationship between historical channel conditions and use it to predict future channel conditions.
Efficacy of this strategy is closely tied to the channel prediction quality but conventional preemptive scheduling mechanism relying solely on the pilot/control signaling and past channel estimates cannot properly handle
unexpected changes caused by the sudden future events
(e.g., blockage or movement direction change).

An aim of this paper is to propose a novel scheduling technique that leverages LMMs to improve the quality of preemptive scheduling. 
In the proposed technique, dubbed as \textit{large multimodal model-based preemptive scheduling} (LMM-PS), 
we aggressively exploit LMMs to predict user's future channel parameters (e.g., distance, angles, and path gain). 
Unlike conventional scheduling techniques that rely exclusively on pilot signals, LMM-PS additionally exploits visual sensing information as an input to LMM to predict the presence of the LoS and strong NLoS paths and geometric information (distance and angles) of users.
In doing so, LMM-PS can quickly understand sudden events in the channel prediction process. 
Since LMM is a powerful tool in extracting correlated features across sequential data and generating subsequent elements~\cite{LMM}, it can be used to predict each user's future channel parameters based on its historical movement captured in the visual sensing information.

The operation of proposed LMM-PS is divided into two major steps (see Fig. 2). 
In the first step, locations and sizes of users and obstacles are extracted via LMM specialized for the object detection (OD), which is designed to detect arbitrary objects specified in the input sentence.
In the second step, both LMM-aided mobility prediction and Transformer-based channel parameter prediction are performed in parallel to predict each user's future LoS and NLoS channel parameters, respectively:
\begin{itemize}
\item \textit{LMM-aided LoS channel parameter prediction}:
To estimate each user's future LoS channel parameters (i.e., azimuth/elevation angle and distance of the LoS path), we leverage an LMM to predict the user's future location based on custom-designed text prompts, which are language instructions derived from sequences of the user's past locations.  
For example, to predict a user’s future location based on its trajectory in the visual sensing information, we use a prompt such as: “\textit{Using the sequence of past x-y coordinate pairs: $(x_{1}, y_{1}) =(930, 380)$, $(x_{2}, y_{2})=(932, 377)$, $(x_{3}, y_{3})=(935, 374)$, predict the next pair $(x_{4}, y_{4})$}". 
Once the user's future location is predicted, the future LoS channel parameters are obtained via a simple coordinate transformation.

\item \textit{Transformer-based NLoS channel parameter prediction}: To estimate the future NLoS channel parameters from the sequence of past channels, we exploit Transformer, a DNN model capturing both long- and short-term correlations between input and output sequences via the attention mechanism~\cite{transformers}.
By assigning larger attention weights
to input sequence (i.e., past channels) that are strongly correlated with the
output values (i.e., future channel parameters), the proposed LMM-PS can extract future NLoS channel parameters and thus predict future channels even in the presence of sudden channel state changes.
\end{itemize}

From the realistic evaluations in dynamically changing 6G environments, we demonstrate that LMM-PS significantly reduces the channel prediction error caused by unexpected events (e.g., blockages) and thus outperforms conventional scheduling approaches in terms of the total system throughput.
For example, we observe that LMM-PS achieves 32\% throughput improvement over the proportional-fair scheduling technique.
Even when compared to the deep reinforcement learning (DRL)-based preemptive scheduling technique, we observe that LMM-PS achieves 11\% throughput gain.

\begin{table}
\centering
\caption{Summary of nomenclature.}
\label{t1}
\begin{tabular}{|c|c|}
\noalign{\smallskip}\noalign{\smallskip}\hline
\textbf{Notation} & \textbf{Description} \\
\hline
BS  & Base station \\
\hline
CU & Central unit \\
\hline
CQI & Channel quality indicator \\
\hline
MCS  & Modulation and coding scheme\\
\hline
SER  &  Symbol error rate\\
\hline
DL & Deep learning \\
\hline
DNN & Deep neural network \\
\hline
RB & Resource block \\
\hline
mmWave  & Millimeter wave \\
\hline
THz & Terahertz \\
\hline
RI & Rank indicator \\
\hline
PMI & Precoding matrix indicator \\
\hline
QoS & Quality-of-service \\
\hline
LoS & Line-of-sight\\
\hline
NLoS & Non-line-of-sight \\ 
\hline
OD & Object detection \\ 
\hline
SoC & System-on-chip \\
\hline
SDM & Spatial division multiplexing  \\
\hline
\end{tabular}
\label{tab:tab001}
\end{table}

We briefly summarize the notations used in this paper. 
We use uppercase and lowercase boldface letters for matrices and vectors, respectively. 
Operations $(\cdot)^{\text{T}}$, $(\cdot)^{\text{H}}$, and $(\cdot)^{^{\dagger}}$ denote transpose, conjugate transpose, and pseudo-inverse, respectively. 
Also, operators $\otimes$ and $\odot$ denote the Hadamard and Kronecker products, respectively. 
$\mathbb{C}$ and $\mathbb{R}$ denote the field of complex numbers and real numbers, respectively. 
Note that the sets of user equipments (UEs), RBs, and MCS indices are denoted as $\mathcal{K} = \lbrace 1, \cdots,  K \rbrace$,  $\mathcal{B} = \lbrace 1, \cdots,  B \rbrace$, and $\mathcal{M} = \lbrace 0, \cdots, M \rbrace$, respectively.
$\lVert \cdot \rVert_{p}$ indicates the $p$-norm, $\mathbf{x}_{i}$ denotes the $i$-th column of matrix $\mathbf{X}$, and $x(i)$ denotes the $i$-th entry of vector $ \mathbf{x}$.
Lastly, $\mathcal{I}$, $\mathcal{I}_{d}$, $\mathcal{I}_{c}$, and $\mathsf{P}$ denote RGB image, depth image, cropped RGB image, and text prompt, respectively.
In~Table \ref{tab:tab001}, we summarize the technical
terms.


The rest of this paper is organized as follows.
In Section II, we discuss the system model and  then formulate the preemptive scheduling problem.
In Section III, we provide a detailed description of  the proposed LMM-PS. 
In Section IV, we present the simulation results and then conclude the paper in Section V.

\section{Preemptive Scheduling Problem}
In this section, we briefly discuss the system model and conventional scheduling techniques. We then formulate the preemptive scheduling problem as a constrained optimization task.

\subsection{Preemptive Scheduling and System Model}\label{sec:2_1}



Main goal of preemptive scheduling is to maximize the total system throughput in the future time slot $T$ by allocating RBs to $K$ UEs and selecting an appropriate MCS index for each UE. 
To indicate the allocation of an RB to each UE at the $t$-th time slot, we define a binary vector $\mathbf{x}^{(t)}_{b} = [ x^{(t)}_{b}(1) \cdots x^{(t)}_{b}(K) ] \in \mathbb{R}^{K}$:
\begin{equation}
  x^{(t)}_{b}(k) =
    \begin{cases}
      1  & \text{if the}\ \text{$b$-th RB is
allocated to the $k$-th UE},\\
      0 & \text{otherwise}.\label{eq:RB_allocation}
    \end{cases}       
\end{equation}
One needs to make sure that each UE needs to use the same MCS index $m$ across all its scheduled RBs~\cite{MCSconstraint}.
To indicate the MCS index assigned to each UE, we define the binary one-hot vector $\mathbf{z}^{(t)}_{k} = [ z^{(t)}_{k}(1) \cdots z^{(t)}_{k}(M) ]  \in \mathbb{R}^{M}$:
\begin{equation}
  z^{(t)}_{k}(m) =
    \begin{cases}
      1  & \text{if}\ \text{MCS $m$ is assigned to the $k$-th UE},\\
      0 & \text{otherwise}.\label{eq:MCS_allocation}
    \end{cases}       
\end{equation}


Based on the 5G NR resource definition in~\cite{5GNRsynchronization}, the downlink throughput $R_{k}$ of the $k$-th UE can be expressed as 
\begin{align}
    R^{(t)}_{k} =  \sum^{M}_{m = 1} \sum^{B}_{b = 1} \Bigg(1 - (1 - (1 - e_{\text{SER},k}^{(t)})^{N_s})\Bigg) \cdot \nonumber \\
    f(z^{(t)}_{k}(m)) x^{(t)}_{b}(k)N_{\text{SDM}}N_{\text{RE}}, 
    \label{eq:throughput}
\end{align}
where $f(\cdot)$ is the MCS order table that maps the selected MCS index $m$ to the unit spectral efficiency (U-SE), which represents the number of bits carried by each resource element (RE), and
$N_{\text{SDM}}$ is the number of spatial division multiplexing (SDM) layers, which is the number of data streams transmitted from the base station (BS) to each UE.
Also, $e_{\text{SER},k}$ is the SER of the $k$-th UE and $N_{\text{RE}}$ is the number of REs contained in each RB.
Note that the total system throughput for $K$ UEs is $R^{(t)}_{\text{total}} = \sum^{K}_{k=1}R^{(t)}_{k}$.



\begin{figure*}[t]
	\centering
    \includegraphics[width=2.0\columnwidth, height=5.0cm]{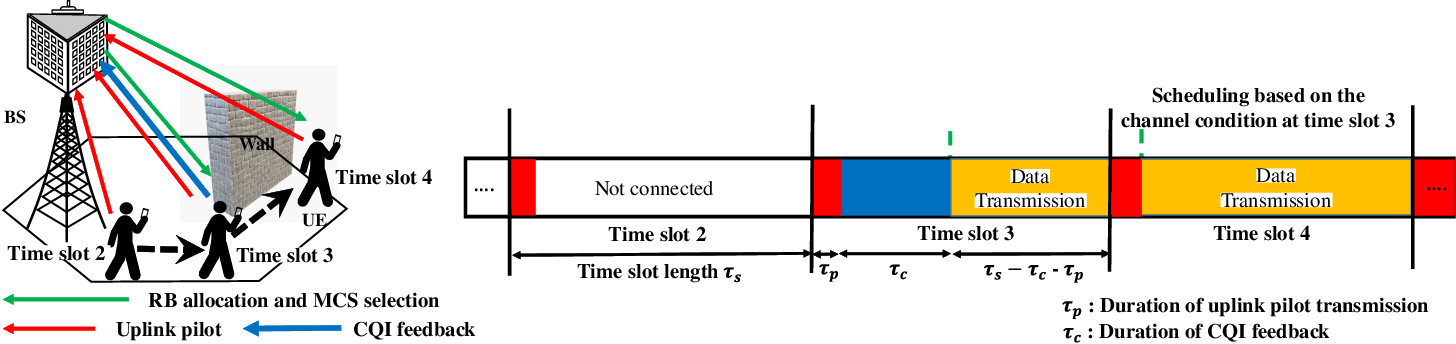}
    \vspace{-0.3cm} 
\caption{Illustration of the conventional CQI feedback-based scheduling. When the BS makes scheduling decisions based on the outdated CQI feedback, they may no longer reflect the true channel condition, leading to degradation of beamforming gain and transmission rate.}
	\label{fig:timeslots}
\end{figure*}
In our work, we consider the multi-input single-output (MISO) system where the BS equipped with a uniform planar array (UPA) of $N_{T} = N_{x} \times  N_{y}$ antennas serves $K$ single-antenna UEs.
In this setup, the received pilot signal $\mathbf{y}^{(t)} \in \mathbb{C}^{N_{T} \times 1}$ from the $k$-th UE is given by 
\begin{equation}
\mathbf{y}^{(t)}_{k} = \mathbf{h}^{(t)}_{k}s + \mathbf{n}^{(t)}_{k},
\end{equation}
where $\mathbf{h}^{(t)}_{k} \in \mathbb{C}^{N_{T} \times 1}$ is the uplink channel vector for the $k$-th UE, $s \in \mathbb{C}$ is the uplink pilot symbol, and $\mathbf{n}^{(t)}_{k} \sim \mathcal{CN}(0, \sigma^2_{n}\mathbf{I})$ is the additive Gaussian noise. 

As for the channel model, we use the block-fading multi-path channel model where the channel varies from block-to-block~\cite{snkim_channelmodel}.
We note here that the proposed technique is an LMM-based technique so that it can flexibly adapt to diverse task objectives and dynamic environments through the training process (e.g., fine-tuning and continual learning).
This means that its performance might not be affected much by changes in the channel environment or interference patterns.
Specifically, the uplink channel vector $\mathbf{h}^{(t)}_{k} \in \mathbb{C}^{M \times 1}$ for the $k$-th UE is
\begin{align}
\mathbf{h}^{(t)}_{k} &= \delta^{(t)}_{k} \mathbf{h}^{(t)}_{\text{LoS}, k} + \mathbf{h}^{(t)}_{\text{NLoS},k} \label{eq:channel} \\
&= 
\delta^{(t)}_{k} \mathbf{h}^{(t)}_{\text{LoS}, k} + 
\sum^{L}_{l = 1}\mathbf{h}^{(t)}_{\text{NLoS}, l, k}, 
\end{align}
where $L$ is the number of NLoS paths and $\delta^{(t)}_k$ represents the status of LoS link of the $k$-th UE:
\begin{equation}
  \delta^{(t)}_{k} =
    \begin{cases}
      1 & \text{if}\ \text{LoS link is available for the $k$-th UE},\\
      0 & \text{else.}\
    \end{cases}       
\end{equation}
Note that the LoS and NLoS channel components are expressed as~\cite{jwson_channelcomponent}
\begin{align}
\mathbf{h}^{(t)}_{\text{LoS}, k} &= \alpha^{(t)}_{\text{LoS}, k}
e^{-j 2\pi f_{c} \frac{r^{(t)}_{\text{LoS}, k}}{c}}
\mathbf{a}(\theta^{(t)}_{\text{LoS},k}, \phi^{(t)}_{\text{LoS}, k}), \label{eq:LoSchannel}\\
\mathbf{h}^{(t)}_{\text{NLoS}, l, k} &= \alpha^{(t)}_{\text{NLoS}, l, k}
e^{-j 2\pi f_{c} \frac{r^{(t)}_{\text{NLoS}, l,k}}{c}}
\mathbf{a}(\theta^{(t)}_{\text{NLoS}, l,k}, \phi^{(t)}_{\text{NLoS}, l,k}),
\label{eq:NLoSchannel}
\end{align}
where $\alpha^{(t)}_{\text{LoS}, k}$ and $\alpha^{(t)}_{\text{NLoS}, l, k}$ are the complex path gains of the LoS path and the $l$-th NLoS path, respectively.
Also,
$c$ is the speed of light, $\mathbf{a}(\theta, \phi) \in \mathbb{C}^{N_T \times 1}$ is the array response vector, $f_{c}$ is the carrier frequency, and $r^{(t)}_{\text{LoS}, k}$ and $r^{(t)}_{\text{NLoS}, l, k}$ are the LoS distance and the distance of the $l$-th NLoS path between the BS and the $k$-th UE, respectively.
Finally,
$(\theta^{(t)}_{\text{LoS},k}, \phi^{(t)}_{\text{LoS}, k})$  and $(\theta^{(t)}_{\text{NLoS}, l,k}, \phi^{(t)}_{\text{NLoS}, l,k})$ are the azimuth/elevation angles of the LoS path and the $l$-th NLoS path, respectively.

\subsection{Conventional Scheduling Techniques}\label{sec:2_2}

In conventional scheduling techniques, UE periodically transmits a CQI value to help the BS select an appropriate MCS for the downlink transmission~\cite{conv4_DL, sJeong}. 
After receiving the transmit pilot signal from the BS, UE computes the downlink channel response and derives the rank indicator (RI) and precoding matrix indicator (PMI)~\cite{conv3_DL}. 
The PMI is mapped to a CQI value using a pre-defined CQI mapping table~\cite{PMItoCQItableandMCSordertable}, using which an MCS index is determined~\cite{CQItoMCStable}. 
Over the years, various deep learning (DL)-based scheduling techniques have been proposed to handle the complex CQI-to-MCS mapping process~\cite{conv2_DL, conv3_DL, conv4_DL}.
Key feature of these techniques is that they use the CQI feedback and channel state information as inputs to DL models.

%
%
%

In dynamically varying 6G environments, performance of conventional scheduling techniques might degrade severely due to densely distributed obstacles, short channel coherence time of mmWave/THz channels, and diverse spectrums.
These factors result in rapid channel variations so that the reliable path (presumably the LoS path) may disappear even with a small UE movement~\cite{YJAhn_SNRdrop, blockageandthreshold}.
Since the CQI feedback is reported at relatively long intervals (typically every 8\,ms~\cite{CQIfeedbackdelayandperiod}), actual channel condition in the current time might differ significantly from the estimated channel condition obtained from the CQI feedback (see Fig. 3).
Clearly, this mismatch will result in a severe degradation in the beamforming gain and transmission rate.

Recently, preemptive scheduling techniques performing the scheduling decision based on predicted future channel conditions have gained much attention~\cite{conv1_preemptive, conv2_preemptive, conv3_preemptive}.
The common wisdom behind these techniques is to capture temporal dependencies in historical CQI measurements to predict future CQI values.
Since these techniques rely exclusively on the past channel information, they often fail to capture sudden channel variations such as blockages, hand tremors, or abrupt changes in movement direction. 
In contrast, by detecting the signal propagation environment using vision sensing images and then inferring sudden events (e.g., signal blockages) from UE movements using LMM, LMM-PS can quickly understand and identify the wireless environments.

\subsection{Preemptive Scheduling Problem Formulation}\label{sec:2_3}



Since the main goal of preemptive scheduling is to maximize the total system throughput in the future time slot $T$, its objective function can be expressed as 
\begin{equation}
    \begin{split}
        R^{(T)}_{\text{total}} = 
        \sum^{K}_{k = 1}\sum^{M}_{m = 1}\sum^{B}_{b = 1}
        \Bigg(1 -  \big(1 - (1 - e_{\text{SER},k}^{(T)})^{N_s}\big)\Bigg) \cdot \\
        f(z^{(T)}_{k}(m)) x^{(T)}_{b}(k)N_{\text{SDM}}N_{\text{RE}}.
    \end{split}
    \label{eq:throughput_maximization}
\end{equation}

Other than the throughput maximization, two major constraints need to be considered: 1) data rate constraint and 2) SER constraint.
First, the rate constraint requires that the total number of RBs allocated to UE $\sum^{B}_{b=1} x^{(T)}_{b}(k)$ needs to be larger than $N^{(T-1)}_{\text{RB},k}$ (the minimum number of RBs to meet the UE's rate requirement $R^{(T-1)}_{k, \text{min}}$ in the $(T-1)$-th slot). 
In the preemptive scheduling, without knowing the exact rate requirements of UEs in the $T$-th (future) slot, the BS needs to allocate RBs for the $T$-th slot at the $(T-1)$-th slot.
To handle this uncertainty, the scheduler ensures that the allocated RBs for the $T$-th slot meet the rate requirements in the $(T-1)$-th slot.


\begin{figure*}[t]
	\centering
    \includegraphics[width=2.0\columnwidth, height=7.0cm]{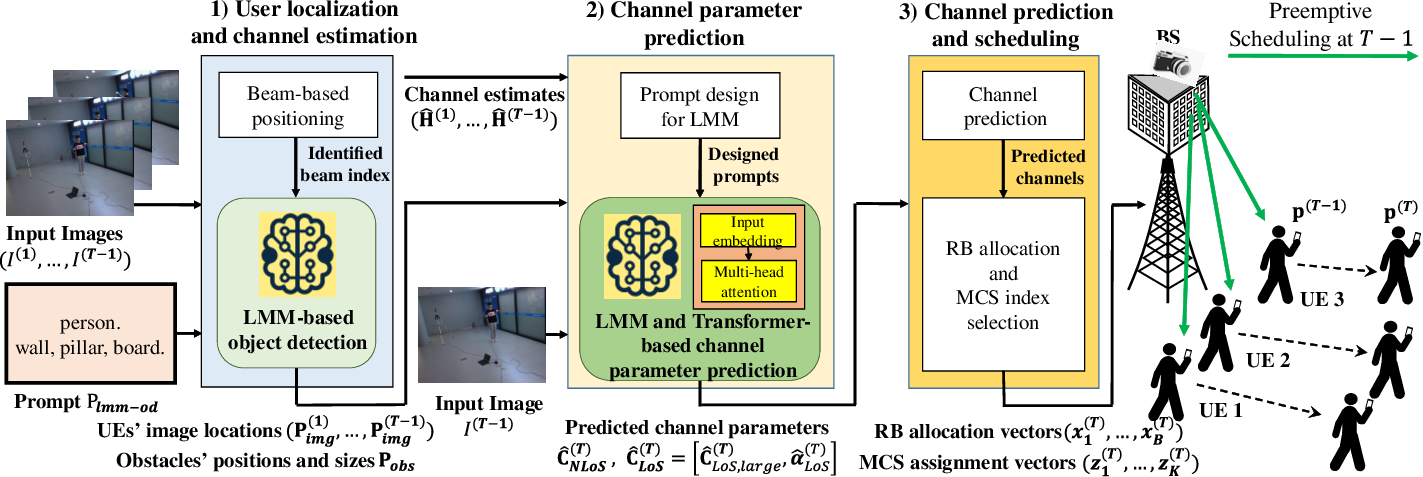}
    \vspace{-0.3cm}
        \caption{Overall process of the proposed LMM-PS.}
	\label{fig:LMM_PS}
    \vspace{-0.4cm}
\end{figure*}
In addition, the SER constraint requires that the SER of the $k$-th UE $e^{(T)}_{\text{SER},k}$ is better (lower) than the target SER $\text{SER}_{k, \text{max}}$~\cite{maximumSNRscheduling}.
To check this, the SER must be computed based on the channel estimate $\hat{\mathbf{h}}_{k}$, which is derived from $\mathbf{y}_{k}$ using the conventional channel estimation technique such as the linear minimum
mean square error (LMMSE).
According to~\cite{SNRtoSER}, the SER is given by
\begin{equation}
  e^{(t)}_{\text{SER}, k} =
    \begin{cases}
      1 - \left(1 - Q\left(\sqrt{ \ \frac{2|\hat{\mathbf{h}}^{(t)}_{k}|^2}{\sigma^2_{n}}}\right)\right)^2 \quad\quad\quad \text{if}\ Q_{m} = 2,\\
     1 - \left( 1 - \frac{2(\sqrt{Q_{m}} - 1)}{\sqrt{Q_{m}}} Q\left( \sqrt{\frac{3 |\hat{\mathbf{h}}^{(t)}_{k}|^2}{(Q_{m} - 1) \sigma^2_{n}}} \right) \right)^2 \,\,\,\, \text{else}, \label{eq:SERfromSNR}
    \end{cases}       
\end{equation}
where $\frac{|\hat{\mathbf{h}}^{(t)}_{k}|^2}{\sigma^2_{n}}$ is the SNR of the $k$-th UE, $Q(\cdot)$ is the Q-function, and $Q_{m}$ is the modulation order chosen from the order table\footnote{For example, if we use the 256QAM order table, then $Q_{m}$ is 2 if the selected MCS index $m_{k} \in \lbrace 0,1,2,3,4 \rbrace$, 4 if  $m_{k} \in \lbrace 5,6,7,8,9,10 \rbrace$, 6 if  $m_{k} \in \lbrace 11,12,13,14,15,16,17,18,19 \rbrace$, 8 if  $m_{k} \in \lbrace  20, 21, 22, 23,24,25,26,27\rbrace$~\cite{PMItoCQItableandMCSordertable}.}.

By incorporating these constraints, the preemptive scheduling problem $\mathcal{P}$ can be formulated as
\begin{subequations}
\begin{align}
     \mathcal{P}: \max_{\lbrace \mathbf{x}^{(T)}_{1}, \cdots, \mathbf{x}^{(T)}_{B},  \mathbf{z}^{(T)}_{1}, \cdots , \mathbf{z}^{(T)}_{K} \rbrace} &\ 
      R^{(T)}_{\text{total}}
    \label{eq:optprob}\\
    \text{s.t.} \quad & 
    \sum^{B}_{b=1} x^{(T)}_{b}(k) \geq N^{(T-1)}_{ \text{RB},k}, \forall k\in \mathcal{K}, \label{eq:datarateconstraint1} \\ 
     & \sum^{K}_{k=1} x^{(T)}_{b}(k) \leq N_{T}, \quad \forall b\in \mathcal{B},\label{eq:resourceconstraint1} \\
    & e^{(T)}_{ \text{SER},k} \leq \text{SER}_{k, \text{max}}, \quad \forall k\in \mathcal{K},\ \label{eq:SERconstraint1} \\
    & \sum^{M}_{m=1} z^{(T)}_{k}(m) = 1. \quad \forall k\in \mathcal{K}. \label{eq:MCS1}  
\end{align}
\end{subequations}
In this work, we assume that 
the number of SDM layers $N_\text{SDM}$ is fixed to 1\footnote{This ensures that $N_\text{SDM}$ does not exceed the number of UE receive antennas, as required in~\cite{SDM}.}.
In our problem $\mathcal{P}$, the resource constraint~\eqref{eq:resourceconstraint1} is employed to ensure that the maximum number of UEs scheduled in an RB does not exceed the number of BS antennas~\cite{SDM}.
Also, we use the MCS assignment constraint~\eqref{eq:MCS1} to guarantee that only one MCS is used across all scheduled RBs for each UE.

By plugging~\eqref{eq:throughput_maximization}, $\mathcal{P}$ can be re-expressed as 
\begin{subequations}
\begin{align}
    \mathcal{P^{'}}: \max_{{\lbrace \mathbf{x}^{(T)}_{1}, \cdots, \mathbf{x}^{(T)}_{B},  \mathbf{z}^{(T)}_{1}, \cdots , \mathbf{z}^{(T)}_{K} \rbrace}} &\ 
     \sum^{K}_{k = 1}\sum^{M}_{m = 1}\sum^{B}_{b = 1}  \nonumber \\ 
     & \bigg(1 -  (1 - (1 - e_{\text{SER},k}^{(T)})^{N_s})\bigg) \cdot \nonumber \\ 
     & \hspace{-10pt} f(z^{(T)}_{k}(m)) x^{(T)}_{b}(k) N_{\text{RE}} \label{eq:optprob_full}\\
     & \hspace{-12pt} \text{s.t.}
    \sum^{B}_{b=1} x^{(T)}_{b}(k) \geq B_{k, \text{min}} , \forall k\in \mathcal{K}, \label{eq:datarateconstraint2} \\ 
     &\ \ \ \hspace{-10pt} \sum^{K}_{k=1} x^{(T)}_{b}(k) \leq N_{T}, \quad  \forall b \in \mathcal{B}, \label{eq:resourceconstraint2} \\
    &\ \ \ \hspace{-10pt} e^{(T)}_{ \text{SER},k} \leq \text{SER}_{k, \text{max}}, \ \forall k\in \mathcal{K}, \label{eq:SERconstraint2} \\
    &\ \ \ \hspace{-10pt} \sum^{M}_{m=1} z^{(T)}_{k}(m) = 1, \quad \forall k\in \mathcal{K}.   \label{eq:MCS2} 
\end{align}
\end{subequations}
Unfortunately, since the objective function of $\mathcal{P^{'}}$ is the total system throughput of the future time slot $T$, the optimal solutions $\lbrace \mathbf{x}^{(T)}_{\text{opt}, b} \rbrace^{B}_{b = 1}$ and $\lbrace \mathbf{z}^{(T)}_{\text{opt}, k} \rbrace^{K}_{k = 1}$  can be obtained only when the UEs' future channels  (i.e., ($\mathbf{h}^{(T)}_{1}, \cdots, \mathbf{h}^{(T)}_{K}$)) are perfectly known, which is obviously unrealistic for the causality issue.
\begin{figure*}[t]
	\centering
        \includegraphics[width=2.0\columnwidth,  height=6cm]{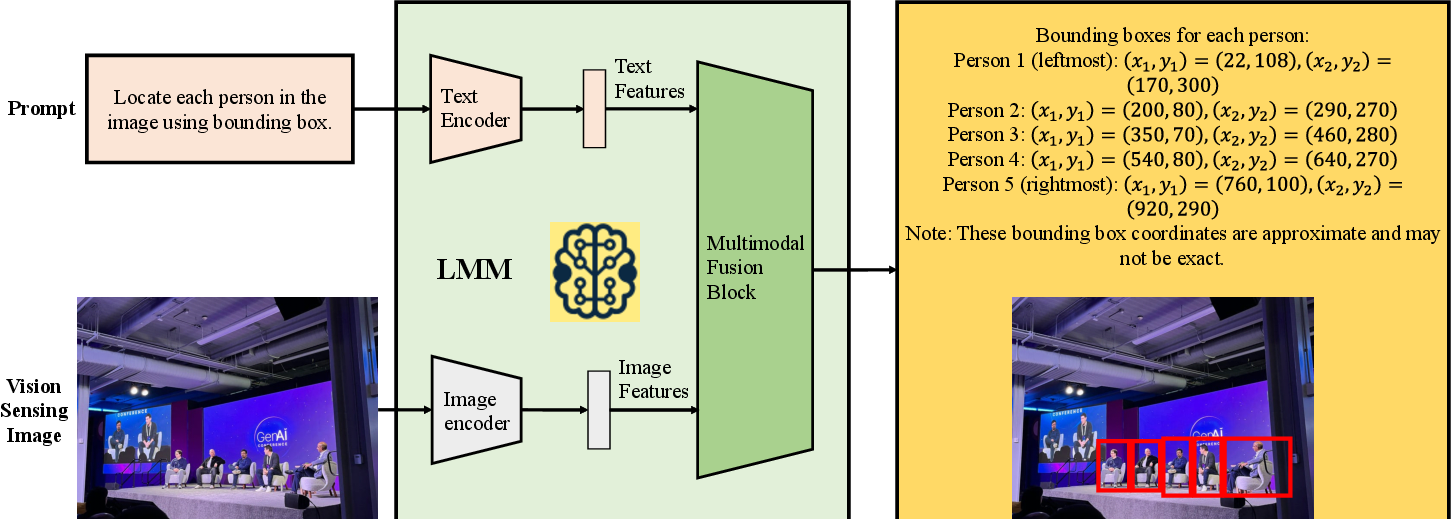}
        \vspace{-0.1cm}
        \caption{Basic structure of LMM.}
        \vspace{-0.5cm}
	\label{fig:basicsofLMM}
\end{figure*}
\section{LMM-based Preemptive Scheduling}

The primary goal of LMM-PS is to solve $\mathcal{P^{'}}$ by predicting the future channels of UEs ($\mathbf{h}^{(T)}_{1}, \cdots, \mathbf{h}^{(T)}_{K}$) with the help of visual sensing and LMMs. 
In a nutshell, LMM-PS predicts the future channels based on sequences of both visual sensing and RF-based past channel estimates. Then, using the predicted future channels,  
LMM-PS finds out $\lbrace \mathbf{x}^{(T)}_{b} \rbrace^{B}_{b = 1}$ and $\lbrace \mathbf{z}^{(T)}_{k} \rbrace^{K}_{k = 1}$ satisfying the constraints~\eqref{eq:datarateconstraint2}-\eqref{eq:MCS2} of $\mathcal{P^{'}}$.

Specifically, LMM-PS consists of three major stages  (see Fig. 4).
First, in the \textit{user localization and channel estimation stage}, the BS uses the LMM-OD to identify UE locations, as well as positions and sizes of obstacles.
For the LMM-OD, we use GLIPv2,  the off-the-shelf LMM trained to detect objects specified in the input sentence~\cite{glipv2}. 
Using the acquired UE locations and uplink pilot signals, the BS performs the channel estimation in each time slot. 

Second, in the \textit{channel parameter prediction stage}, the BS predicts each UE's future channel parameters (e.g., distance, angles, and path gain).
Since the channel in mmWave/THz environments is modeled by a small number of propagation paths (LoS and $1\sim4$ NLoS paths) and each path consists of few channel parameters, LMM-PS can effectively predict the future channel by estimating these parameters instead of the full-dimensional channel $\mathbf{h}^{(T)}_{k}$.
To predict UE's future large-scale LoS channel parameters (distance/angles of the LoS path) from its past image locations, the BS uses LLaVA, an LMM capable of capturing spatio-temporal correlations across multimodal sequential data by combining visual and textual information.
 Since LLaVA uses both image and text inputs to generate the next elements, it can effectively predict UE's future channel parameters based on its past trajectory~\cite{LLaVA}.  Moreover, as a language-based model, LLaVA can flexibly adapt to new task objectives (e.g., changing the target UE types) using the text prompt.
 LLaVA can also readily adapt to various deployment environments through a relatively simple fine-tuning process.
 For example, using the low-rank adaptation (LoRA)-based fine-tuning technique~\cite{lora}, we can train only the parameters of the newly added block (called adapter), which helps LLaVA to quickly perform robust UE trajectory 
 prediction across various deployment scenarios.

Simultaneously, the BS employs Transformers to predict UE's future NLoS channel parameters (and LoS path gain) from past channel estimates.
Since Transformer assigns relatively large attention weights to input data (i.e., past channels) that are highly correlated to
 the output values (i.e., future channel parameters), it can effectively capture
 long-term spatial correlation that tends to be preserved even under sudden environmental changes (e.g., appearance of new scatterers), resulting in an enhancement of the channel prediction accuracy.

Finally, in the \textit{channel prediction and scheduling stage}, LMM-PS solves the problem $\mathcal{P^{'}}$ in~\eqref{eq:optprob_full} using the predicted future channels.
To be specific, the BS generates future channel estimates by substituting the acquired channel parameters into the channel models \eqref{eq:LoSchannel} and \eqref{eq:NLoSchannel}.
These estimates are then used to compute the proportional-fair (PF) metric~\cite{PFmetric} determining the UE's scheduling priority.
Then, among the RBs satisfying the constraints~\eqref{eq:datarateconstraint2}-\eqref{eq:resourceconstraint2}, the BS allocates the maximum number of RBs to UEs with the highest PF values, while UEs with smaller PF values are allocated fewer RBs. 
Similarly, among the MCS indices satisfying the constraints~\eqref{eq:SERconstraint2}-\eqref{eq:MCS2}, the highest eligible MCS index is chosen for each UE.

\subsection{Role of LMMs in Channel  Prediction}\label{sec_3_0}
\begin{figure*}[t]
	\centering
        \includegraphics[width=2.0\columnwidth,  height=7.0cm]{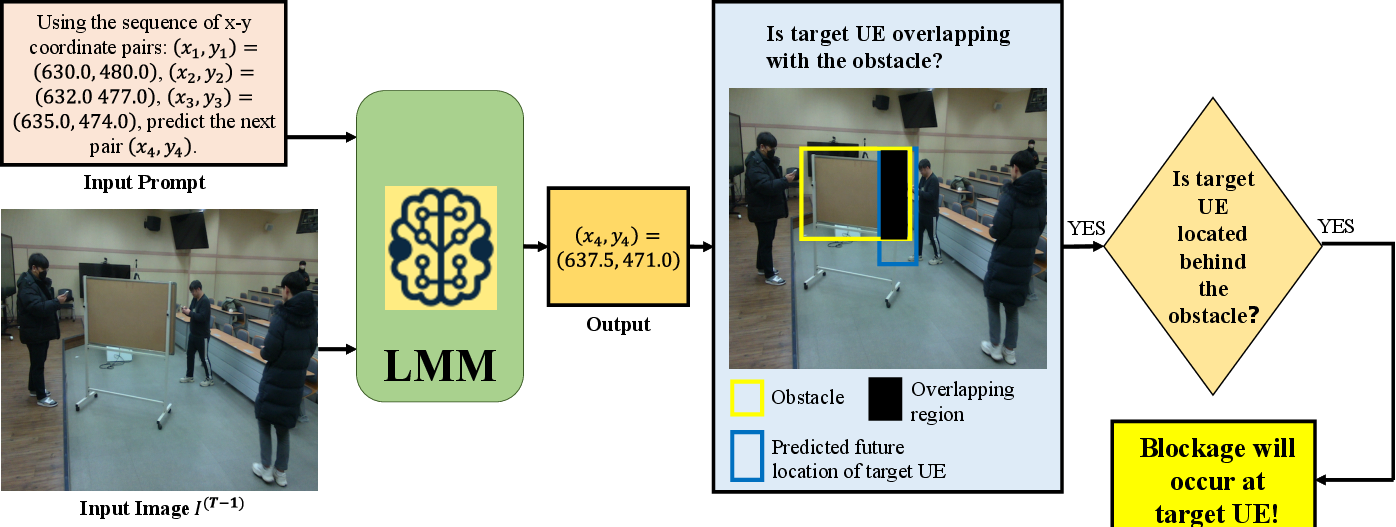}
        \vspace{-0.5cm}
        \caption{Illustration of LMM-aided blockage prediction.}
        \vspace{-0.5cm}
	\label{fig:blockage}
\end{figure*}
LMMs are extensions of LLMs designed to process inputs from multiple modalities (e.g., images and text). Key components of LMMs include pre-trained encoders for each modality (e.g., ViT~\cite{vit} for images and BERT~\cite{bert} for text) and a fusion network that integrates features from each modality to generate task-specific predictions (see Fig.~\ref{fig:basicsofLMM})~\cite{structure_of_LMMs}. 

The major benefits of using LMM in LMM-PS when compared to the conventional DNN can be summarized as:
\begin{itemize}
\item  In contrast to the conventional object detector (OD) that can only identify objects of pre-defined classes (e.g., cats and laptops), the LMM tailored for object detection (LMM-OD) can readily detect arbitrary objects, both inside and outside of these classes.
This is because while conventional OD relies solely on image inputs, LMM-OD exploits both image and text inputs to understand an input sentence and identify the image locations of the objects mentioned in the sentence.
This means that LMM-OD can flexibly adapt to new task objectives or environmental changes.
For example, if obstacles causing LoS blockages (e.g., robots and motorcycles) newly appear in the captured image, one can simply update the prompt so that LMM-OD is not distracted and focuses only on the target object.

\item Unlike the conventional trajectory prediction techniques relying solely on past UE trajectories, LMMs can capture the spatio-temporal relationships in the scene to predict UEs' future pixel-level locations.
Using the improved prediction capability, one can detect unexpected events (e.g., LoS blockages or sudden changes in movement direction) and then consider them in the channel prediction process.
For example, LMMs can identify UEs that may overlap with obstacles by comparing their predicted locations with the obstacles' positions and sizes (see Fig.~\ref{fig:blockage}), which cannot be achieved in conventional DNN technique. Note that the high-precision LoS blockage prediction enabled by LMMs is crucial in 6G channel environments, where the ultra-high frequency spectrum (e.g., FR2 band and FR3 mid-upper band) is explored. In such environments, channel becomes LoS dominant due to the severe signal attenuation in NLoS paths (at most a few paths will survive),  which implies that even a single LoS blockage misdetection might severely degrade the channel prediction accuracy.

\item Updating weights of DNNs whenever the environment or task changes is not practical for real-world scenarios.
In contrast, since LMMs are language-based models, one can easily update/modify the task without the model training using the prompt engineering. 
Among various prompting techniques, chain-of-thought (CoT) prompting, which guides LMMs to reason through intermediate steps, is used in our study.  
For example, when a prompt like “\textit{Using the sequence of past x-y coordinate pairs: $(x_{1}, y_{1}) =(630, 480)$, $(x_{2}, y_{2})=(632, 477)$, $(x_{3}, y_{3})=(635, 474)$, predict the next pair $(x_{4}, y_{4})$}” is given, and if the model generates an implausible output that deviates from the established pattern (e.g., $(x_{4}, y_{4}) = (941, 611)$), then additional prompts can be used to enhance the CoT reasoning (e.g., “\textit{If the predicted pair is physically implausible, predict the next pair $(x_{6}, y_{6})$ from the longer sequence: $(x_{1}, y_{1}) =(625, 485)$, $(x_{2}, y_{2}) =(628, 483)$, $(x_{3}, y_{3}) =(630, 480)$, $(x_{4}, y_{4})=(632, 477)$, $(x_{5}, y_{5})=(635, 474)$}”).
\end{itemize}

\subsection{LMM-based User Localization and Channel Estimation}\label{sec_3_1}

The first stage of LMM-PS, called the \textit{user localization and channel estimation} stage, is divided into four main steps: 1) beam-based positioning, 2) object detection in RGB image, 3) multi-user ID identification, and 4) LoS/NLoS channel components estimation.

\subsubsection{Beam-based Positioning}
In this step, we transmit a few beams to identify the approximate direction of the UEs. 
For the identified physical area, we use the RGB-d camera installed at the BS to capture both RGB and depth images.

\subsubsection{Object Detection in RGB Image}
From the captured RGB image $\mathcal{I}^{(t)}$, we use the LMM-OD $g_{\text{lmm-od}}$ to identify the locations of UEs, as well as the positions and sizes of obstacles.

\begin{figure*}[t]
	\centering
    \includegraphics[width=2.0\columnwidth,  height=8.5cm]{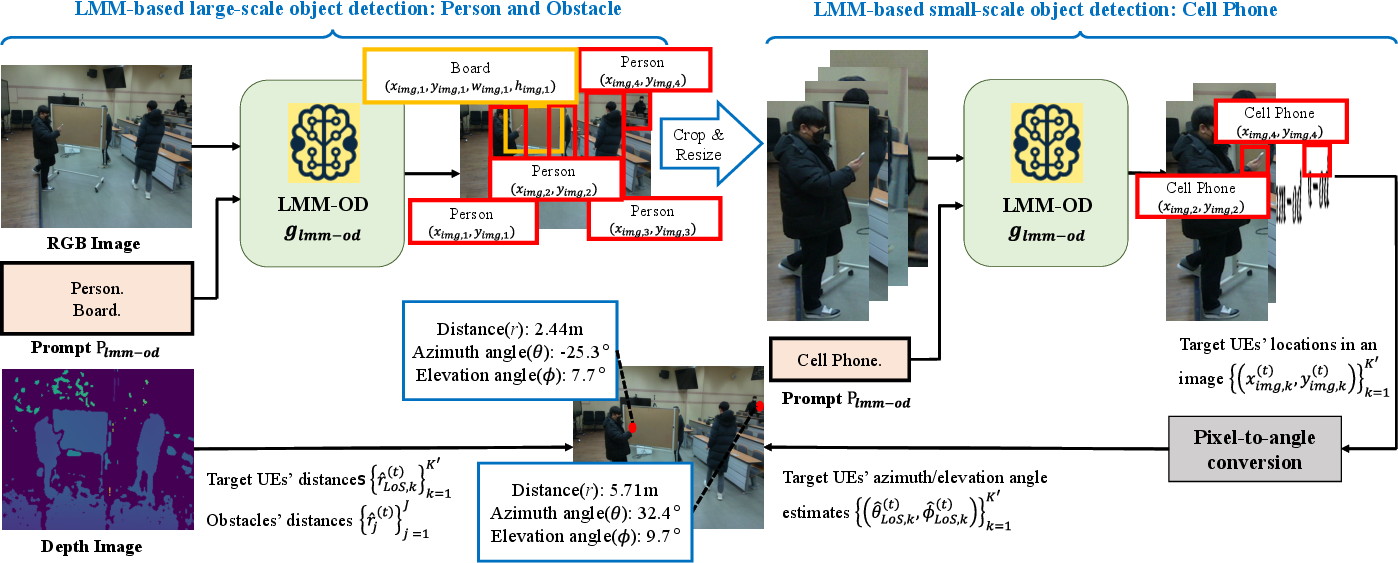}
    \vspace{-0.3cm}
	\caption{Illustration of the sensing and LMM-aided user localization and obstacle identification.}
	\vspace{-0.2cm}
	\label{fig:languageguidedobjectdetection}
\end{figure*}

\begin{itemize}
\item \textbf{User Localization}: 
Since most UEs are small-scale mobile devices (e.g., cell phones and laptops)~\cite{vomtc, beamforming}, they occupy only a few pixels in $\mathcal{I}^{(t)}$, making it difficult for $g_{\text{lmm-od}}$ to identify them.  
To overcome this shortcoming, we adopt a two-stage object detection process where $g_{\text{lmm-od}}$ identifies persons holding the UEs in $\mathcal{I}^{(t)}$ and then detects each UE in the cropped image $\mathcal{I}^{(t)}_{c}$ of the identified person.  
For example, $g_{\text{lmm-od}}$ identifies persons in $\mathcal{I}^{(t)}$ using the text prompt “\textit{person}" and $\mathcal{I}^{(t)}$ as inputs.
Then, $g_{\text{lmm-od}}$ uses $\mathcal{I}^{(t)}_{c}$ and the text prompt “\textit{cell phone}" as inputs to identify the image locations of $K^{'}$ cell phones 
$\mathbf{P}^{(t)}_{\text{img}} = \lbrace (x^{(t)}_{\text{img},k},y^{(t)}_{\text{img},k}) \rbrace^{K^{'}}_{k = 1}$.
By combining the angles of each UE $(\hat{\theta}^{(t)}_{\text{LoS},k}, \hat{\phi}^{(t)}_{\text{LoS},k})$ obtained via pixel-to-angle conversion\footnote{Using the focal length of the RGB-d camera $d_{f}$ (e.g., 50\,mm), the camera's direction angles $(\theta_{i}, \phi_{i})$, and the 2D location $(x_{\text{img}},y_{\text{img}})$ of each detected object, pixel-to-angle conversion can be performed as in~\cite{YJAhn_SNRdrop}.}, along with the distance $\hat{r}^{(t)}_{\text{LoS},k}$ measured from the captured depth image $\mathcal{I}^{(t)}_{d}$, we obtain $K^{'}$ sets of LoS channel parameter 
estimates (see Fig.~\ref{fig:languageguidedobjectdetection}).

\item \textbf{Obstacle Identification}: 
 To determine the locations and sizes of obstacles in $\mathcal{I}^{(t)}$, $g_{\text{lmm-od}}$ uses a text prompt specifying the types of obstacles to identify.
 For instance, using the input text prompt “\textit{wall, board}", $g_{\text{lmm-od}}$ returns $\mathbf{P}_{\text{obs}} = \lbrace (x_{\text{img},j},y_{\text{img},j}, w_{\text{img},j}, h_{\text{img},j}) \rbrace^{J}_{j = 1}$, which contains the center coordinates, width, and height of walls and boards in $\mathcal{I}^{(t)}$.


\end{itemize}

\subsubsection{Multi-user ID identification}
\begin{figure*}[t]
	\centering
        \includegraphics[width=2.0\columnwidth,  height=7.0cm]{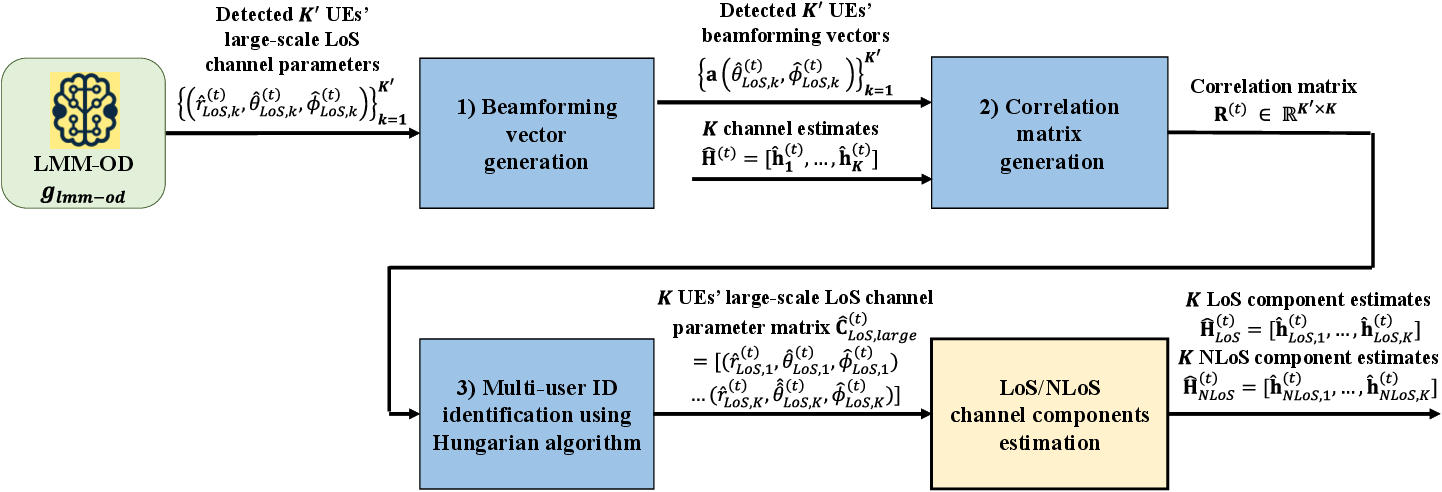}
        \vspace{-0.1cm}
        \caption{Step-by-step operations of multi-user identification process in LMM-PS.}
        \vspace{-0.5cm}
	\label{fig:operationsLMMMUS}
\end{figure*}

Since LMM-OD detects multiple UEs in the captured image $\mathcal{I}^{(t)}$, each detected UE needs to be identified by its assigned ID from the BS.
To this end, we perform the multi-user ID identification through the following steps (see Fig.~\ref{fig:operationsLMMMUS}):

\begin{enumerate}
    \item After acquiring the azimuth and elevation angles $(\hat{\theta}^{(t)}_{\text{LoS},k}, \hat{\phi}^{(t)}_{\text{LoS},k})$ of $K'$ detected UEs, we generate a beamforming vector for each detected UE: $\mathbf{f}^{(t)}_{k} = \mathbf{a}(\hat{\theta}^{(t)}_{\text{LoS},k}, \hat{\phi}^{(t)}_{\text{LoS},k})$.
    \item Since the channel estimates for all $K$ UEs (served by the BS) are available via the conventional channel estimation technique, we identify the ID of each detected UE by determining which channel estimate $\mathbf{\hat{h}}^{(t)}_{k}$ maximizes the correlation with UE's beamforming vector $\mathbf{f}^{(t)}_{k}$. 
    For this purpose, we compute the correlation $| \mathbf{\hat{h}}^{(t) \text{H}}_{k} \mathbf{f}^{(t)}_{k}|$ between each of the $K'$ beamforming vectors and the $K$ channel estimates, generating a correlation matrix $\mathbf{R}^{(t)} \in \mathbb{R}^{K' \times K}$. 
    \item 
    To correctly identify the IDs of all detected UEs, accurate one-to-one matching between the $K'$ beamforming vectors and the $K$ channel estimates is crucial. To this end, we formulate the matching problem using $\mathbf{R}^{(t)}$, where the cost of assigning a beamforming vector $\mathbf{f}^{(t)}_{k}$ to a channel estimate $\mathbf{\hat{h}}^{(t)}_{k}$ is defined as the inverse of the correlation $| \mathbf{\hat{h}}^{(t) \text{H}}_{k} \mathbf{f}^{(t)}_{k}|$. We then solve this problem using the Hungarian algorithm~\cite{Hungarian}, which computes the optimal one-to-one matching between the two sets by iteratively minimizing the total matching cost.
    In our case, we perform one-to-one matching between beamforming vectors and channel estimates.

\end{enumerate}
For any UE not detected in $\mathcal{I}^{(t)}$, we select the beamforming vector (i.e., beam codeword) that maximizes the reference signal received power (RSRP) from the beam codebook. Using the chosen beam codeword index, we extract rough estimates of the UE's distance and angles~\cite{codebook_based_paramacquisition}. 
After identifying the IDs of all $K'$ detected UEs, we determine the large-scale LoS channel parameters for all 
$K$ UEs by matching each set of distances and angles estimated using LMM-OD with the identified UE: $\hat{\mathbf{C}}^{(t)}_{\text{LoS, large}} = [ (\hat{r}^{(t)}_{\text{LoS},1}, \hat{\theta}^{(t)}_{\text{LoS},1}, \hat{\phi}^{(t)}_{\text{LoS},1}) \cdots  (\hat{r}^{(t)}_{\text{LoS},K}, \hat{\theta}^{(t)}_{\text{LoS},K}, \hat{\phi}^{(t)}_{\text{LoS},K})]$.

\subsubsection{LoS/NLoS Channel Components Estimation}
Once the large-scale LoS channel parameter estimates are acquired, we estimate the small-scale LoS path gain parameter from the received pilot signal $\mathbf{y}^{(t)}_{k}$ using the least squares (LS) technique:
\begin{equation}
\hat{\alpha}^{(t)}_{\text{LoS}, k} = e^{j 2\pi f_{c} \frac{\hat{r}^{(t)}_{\text{LoS},k}}{c}}(\mathbf{a}(\hat{\theta}^{(t)}_{\text{LoS},k}, \hat{\phi}^{(t)}_{\text{LoS}, k}))^{\dagger}s \mathbf{y}^{(t)}_{k}.
\end{equation} 
Using both the acquired large-scale and small-scale channel parameters, we estimate the full-dimensional LoS channel component for each UE as $\hat{\mathbf{h}}^{(t)}_{\text{LoS}, k} = 
\hat{\alpha}^{(t)}_{\text{LoS}, k}
e^{-j 2\pi f_{c} \frac{\hat{r}^{(t)}_{\text{LoS},k}}{c}}\mathbf{a}(\hat{\theta}^{(t)}_{\text{LoS},k}, \hat{\phi}^{(t)}_{\text{LoS}, k})$.

Since the channel $\mathbf{h}^{(t)}_{k}$ consists of LoS and NLoS components (see~\eqref{eq:channel}), we estimate the NLoS component $\hat{\mathbf{h}}^{(t)}_{\text{NLoS}, k}$ of each visible UE\footnote{By visible and non-visible UEs, we mean UEs that are present and absent in the captured image $\mathcal{I}^{(t)}$, respectively.} by subtracting the LoS component estimate $\hat{\mathbf{h}}^{(t)}_{\text{LoS}, k}$ from the channel estimate $\mathbf{\hat{h}}^{(t)}_{k}$ (i.e., $\hat{\mathbf{h}}^{(t)}_{\text{NLoS}, k} = \hat{\mathbf{h}}^{(t)}_{k} - \hat{\mathbf{h}}^{(t)}_{\text{LoS}, k}$).
For non-visible UEs, we can readily assume that their LoS paths are blocked by obstacles so that their channels consist solely of NLoS components, setting $\hat{\mathbf{h}}^{(t)}_{\text{LoS}, k} = \mathbf{0}$ and $\hat{\mathbf{h}}^{(t)}_{\text{NLoS}, k} = \hat{\mathbf{h}}^{(t)}_{k}$.

\subsection{Channel Parameter Prediction Using LMM and Transformers}\label{sec_3_2}

In this subsection, we discuss the second stage of LMM-PS, referred to as the \textit{channel parameter prediction} stage. 
To predict future large-scale LoS channel parameters based on the spatio-temporal relationships in the UEs' past image locations, LMM-PS uses LLaVA, the LMM capable of capturing these relationships and generating the next elements. 
LMM-PS also employs Transformers to predict future small-scale LoS path gain and NLoS channel parameters from past LoS/NLoS components. By assigning higher attention to past channel estimates strongly correlated with future parameters, Transformers can predict these values effectively.


\subsubsection{
Large-Scale LoS Channel Parameter Prediction Using LMM}

\begin{figure*}[t]
	\centering
        \includegraphics[width=2.0\columnwidth,  height=5.5cm]{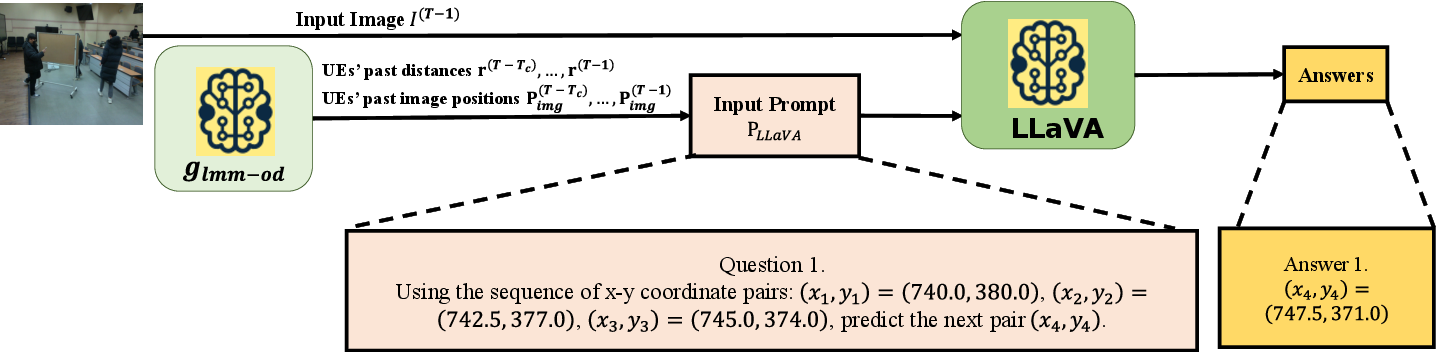}
        \vspace{-1.0cm}
        \caption{Prompt Design.}
        \vspace{-0.1cm}
	\label{fig:promptgeneration}
\end{figure*}

Since LLaVA generates future channel parameter predictions based on the input text prompt $\mathsf{P}_{\text{LLaVA}}$, designing $\mathsf{P}_{\text{LLaVA}}$ carefully is crucial for clear and accurate guidance.  
Specifically, $\mathsf{P}_{\text{LLaVA}}$ consists of two types of prompts: one for predicting UEs' future angles and another for their distances.  
\begin{itemize}
\item \textbf{Future UE angles prediction prompt}: 
To predict each UE's future azimuth/elevation angles $ (\hat{\theta}^{(T)}_{\text{LoS},k}, \hat{\phi}^{(T)}_{\text{LoS},k})$ from its past image locations $\lbrace (x^{(t)}_{\text{img},k}, y^{(t)}_{\text{img},k})\rbrace_{t = T-T_{c}}^{T-1}$, we use a prompt such as: “\textit{Using the sequence of past x-y coordinate pairs: $(x_{1}, y_{1}) = (740.0, 380.0)$, $(x_{2}, y_{2}) = (742.5, 377.0)$, $(x_{3}, y_{3}) = (745.0, 374.0)$, predict the next pair $(x_{4}, y_{4})$}”.
Once LLaVA predicts the next image location $(x^{(T)}_{\text{img},k},y^{(T)}_{\text{img},k})$, it is converted into angle estimates $(\hat{\theta}^{(T)}_{\text{LoS},k},\hat{\phi}^{(T)}_{\text{LoS},k})$ via pixel-to-angle conversion.  
\item \textbf{Future UE distance prediction prompt}: To predict the future distance of each UE from the BS $\hat{r}^{(T)}_{\text{LoS},k}$, we use a sequence of past measured distances $\lbrace \hat{r}^{(t)}_{\text{LoS},k} \rbrace_{t = T-T_{c}}^{T-1}$.  
For example, we use a prompt such as: “\textit{Using the sequence of past distance values: $r_{1} = 12.3$, $r_{2} = 11.2$, $r_{3} = 10.3$, predict the next distance value $r_{4}$}" (see Fig.~\ref{fig:promptgeneration}).
\end{itemize}


Once LLaVA predicts future UE image locations and distances, LMM-PS leverages this information to detect sudden events.  
Specifically, to identify potential future LoS blockages, LMM-PS checks whether each UE's predicted image location overlaps with any obstacle in $\mathcal{I}^{(T-1)}$ by comparing it against the obstacles' positions and sizes.  
If an overlap is detected, then LMM-PS determines the future LoS link status of the UE  $\hat{\delta}^{(T)}_{k}$ by checking whether its predicted distance from the BS $\hat{r}^{(T)}_{\text{LoS},k}$ is shorter than the overlapping obstacle's distance from the BS $\hat{r}_{j}$:
\begin{equation}
  \hat{\delta}^{(T)}_{k} =
    \begin{cases}
      1 & \text{if}\ \hat{r}^{(T)}_{\text{LoS},k} < \hat{r}_{j}, \forall (k, j) \in {\mathcal{O}^{(T)}}\\
      0 & \text{else,}\
    \end{cases} 
    \label{eq:blockageprediction}
\end{equation}
where $\mathcal{O}^{(T)}$ is the index set of UEs and obstacles whose predicted image locations overlap. 

\subsubsection{LoS Path Gain and NLoS Channel Parameter Prediction Using Transformers}
Since LoS and NLoS channel components depend on their respective channel parameters, we design two separate Transformer-based channel predictors: one for LoS (TCP-L) and the other for NLoS (TCP-N), to effectively predict each UE's future LoS path gain and NLoS channel parameters.

In TCP-L and TCP-N, we use past estimated LoS and NLoS channel components to construct the input matrices $\widetilde{\mathbf{H}}_{\text{LoS}}$ and $\widetilde{\mathbf{H}}_{\text{NLoS}}$, respectively. These matrices are embedded into $\mathbf{S}_{\text{embed}} \in \mathbb{R}^{K(T-1) \times N{T}}$ and then passed through $N_{\text{attn}}$ attention blocks, where each block sequentially applies multi-head attention, layer normalization, a fully-connected (FC) layer, and another layer normalization.
This process generates 
the feature matrices $\mathbf{G}_{\text{LoS}, B} \in \mathbb{R}^{K(T-1) \times N_{T}}$ and $\mathbf{G}_{\text{NLoS}, B} \in \mathbb{R}^{K(T-1) \times N_{T}}$ for TCP-L and TCP-N, which are converted to the UEs' future LoS path gains
$\hat{\bm{\alpha}}^{(T)}_{\text{LoS}} \in \mathbb{R}^{ K \times 2}$ and their initially predicted NLoS channel parameters $\hat{\mathbf{C}}^{(T - 1)}_{\text{NLoS}} \in \mathbb{R}^{ K \times 5L_{\text{max}}}$ using multiple FC layers, respectively\footnote{Note that $5$ of $\hat{\mathbf{C}}^{(T)}_{\text{NLoS}}$ comes from real and imaginary values of path gains, azimuth and elevation angles, and distance.} (see Fig. 10).

To handle the variation in the number of NLoS paths $L$ caused by sudden scatterers, TCP-N exploits visual sensing and Transformer.
Specifically, TCP-N extracts the image feature matrix $\mathbf{F} \in \mathbb{R}^{D \times 5L_{\text{max}}}$ from the captured image $\mathcal{I}^{(T-1)}$ using an image feature extractor\footnote{
In this work, we use the Swin Transformer-based image feature extractor~\cite{swin} to mitigate the complexity overhead.
}, while embedding $\hat{\mathbf{C}}^{(T - 1)}_{\text{NLoS}}$ into $\mathbf{Z}_{\text{embed}, F} \in \mathbb{R}^{K \times 5L_{\text{max}}}$. Next, $\mathbf{F}$ and $\mathbf{Z}_{\text{embed}, F}$ are passed through $N_{\text{attn}}$ attention blocks to generate the feature matrix $\mathbf{G}_{\text{NLoS}, F} \in \mathbb{R}^{K \times 5L_{\text{max}}}$, which is then converted to two outputs using multiple FC layers: (1) future NLoS channel parameter predictions $\hat{\mathbf{C}}^{(T)}_{\text{NLoS}} \in \mathbb{R}^{ K \times 5L_{\text{max}}}$ and (2) existence probabilities (i.e., score values $\hat{\mathbf{S}}^{(T)}_{\text{NLoS}} \in \mathbb{R}^{ K \times L_{\text{max}}}$) of NLoS paths.
By assigning relatively large attention weights to input data that are highly correlated to the NLoS channel parameters (i.e., RGB pixels associated with path scatterers), TCP-N can extract the channel-related visual features from the visual sensing image and then predict the channel parameters.
Finally, by selecting the set of predicted channel parameters whose score value is greater than a pre-defined
threshold, TCP-N can flexibly adjust the number of predicted NLoS paths.

\begin{figure*}[t]
	\centering
	\includegraphics[width=2.0\columnwidth, height=11cm]{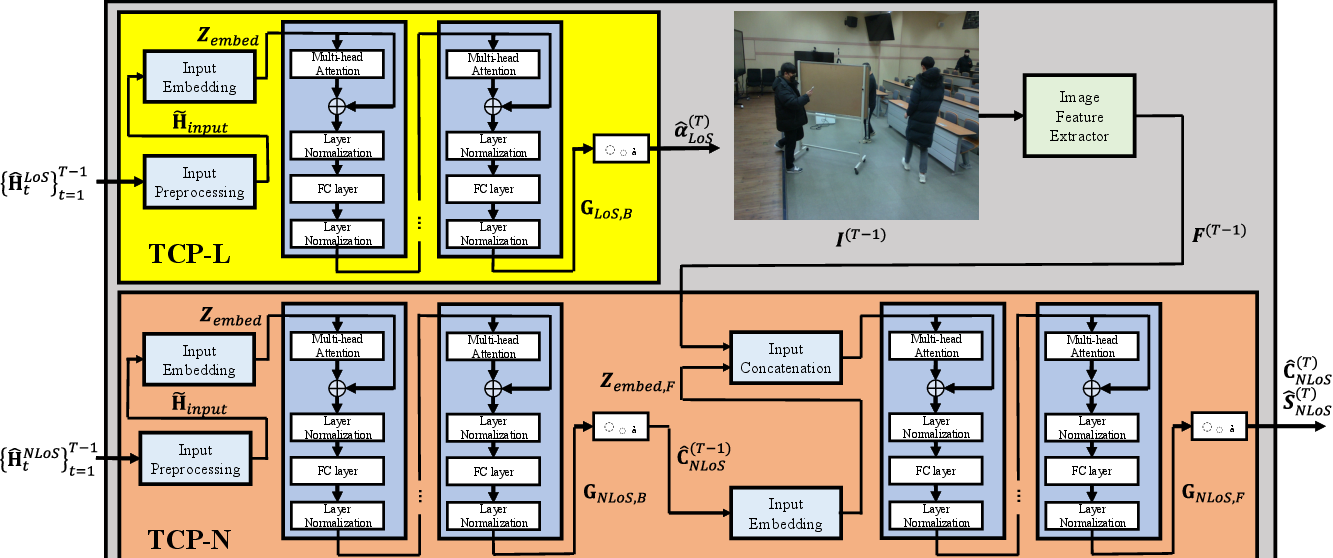}
     \vspace{-0.3cm}
	\caption{Detailed architecture of TCP in LMM-PS.}
 \vspace{-0.5cm}
	\label{fig:TCP}
\end{figure*}

Once the channel prediction using the acquired future channel parameters is completed, we compute the PF metric for all UEs, which determines their scheduling priority. The BS then allocates the maximum number of RBs to UEs with the highest PF values, while allocating fewer RBs to UEs with smaller PF values, subject to the constraints~\eqref{eq:datarateconstraint2}-\eqref{eq:resourceconstraint2}. Similarly, the highest MCS index is chosen from the set of indices satisfying the constraints~\eqref{eq:SERconstraint2}-\eqref{eq:MCS2}.

\subsection{TCP Loss Function Design and Training Process}\label{sec_3_4}

In the training phase, we optimize the network parameters of TCP-L and TCP-N by minimizing the regression loss function $L$, which quantifies the difference between the true and predicted channel parameters. 
To obtain the ground-truth channel parameters, we use MATLAB's ray-tracing function. In this process, we can identify signal propagation paths and their associated channel parameters, and then apply a realistic path loss model to generate synthetic datasets.
The loss functions for TCP-L and TCP-N are expressed as
\begin{align}
L_{\text{TCP-L}} &= \Vert  \bm{\alpha}_{\bigtriangleup}  \Vert^{2}_{2} + \lambda\Vert  \bm{\alpha}_{\bigtriangleup}  \Vert_{1}, \nonumber\\
 \bm{\alpha}_{\bigtriangleup} &=  \bm{\alpha}^{(T)}_{\text{LoS}}  -  \hat{\bm{\alpha}}^{(T)}_{\text{LoS}}, 
 \end{align}
 \begin{align}
L_{\text{TCP-N}} &= \Vert  \mathbf{C}_{\bigtriangleup}  \Vert^{2}_{2} + \lambda \Vert  \mathbf{C}_{\bigtriangleup}  \Vert_{1}, \nonumber\\
 \mathbf{C}_{\bigtriangleup} &=  \mathbf{C}^{(T)}_{\text{NLoS}}  -  \hat{\mathbf{C}}^{(T)}_{\text{NLoS}}, 
 \end{align}
where $\Vert \bm{\alpha}_{\bigtriangleup} \Vert^{2}_{2}$ and $\Vert \mathbf{C}_{\bigtriangleup} \Vert^{2}_{2}$ are the mean square error (MSE) losses,  $\Vert \bm{\alpha}_{\bigtriangleup} \Vert_{1}$ and $\Vert \mathbf{C}_{\bigtriangleup} \Vert_{1}$ are the mean absolute error (MAE) losses, and $\lambda$ is the regularization weight for the MAE loss term.
Note that in our loss function design, we combine both MAE and MSE to mitigate the vanishing gradient problem.
When the MSEs $\Vert \bm{\alpha}_{\bigtriangleup} \Vert^{2}_{2}$ and $\Vert \mathbf{C}_{\bigtriangleup} \Vert^{2}_{2}$ approach zero, their gradients also become very small, which can slow down the convergence of TCP. 
 By including the MAEs $\Vert \bm{\alpha}_{\bigtriangleup} \Vert_{1}$ and $\Vert \mathbf{C}_{\bigtriangleup} \Vert_{1}$, which have constant gradients, we can achieve more stable and consistent training~\cite{YJAhn_SNRdrop}.

\section{Experiments and Discussions}

\subsection{Simulation Setup}\label{sec_5_1}
\begin{table}
\centering
\caption{Simulation parameters}
\label{t1}
\begin{tabular}{|c|c|}
\noalign{\smallskip}\noalign{\smallskip}\hline
\textbf{Parameters} & \textbf{Values} \\
\hline
Carrier frequency $f_{c}$ & $28\,$GHz\\
\hline
Number of BS antennas $N_{T} $ & 64 \\
\hline
Time slot length $\tau_{s}$ & $100\,$ms \\
\hline
Bandwidth $W$  &  $100\,$MHz\\
\hline
Number of symbols in RB $N_s$ & $14$ \\
\hline
Minimum user height $h_{\text{min}}$ & $0.5\,$m \\
\hline
Maximum user height $h_{\text{max}}$ & $2\,$m \\
\hline
Maximum user speed $v_{\text{max}}$ & $25\,$km/h\\
\hline
Maximum number of NLoS paths $L_{\text{max}}$ & $3$ \\
\hline
Number of training samples & $25,600$ \\
\hline
Number of validation samples & $3,000$ \\
\hline
Number of test samples & $7,400$ \\
\hline
Number of hidden units $E$ & $64$\\
\hline
Number of attention blocks $N_{\text{attn}}$ & $5$ \\ 
\hline
Number of training epochs & $500$\\ 
\hline
Regularization coefficient $\lambda$ & $1.0$  \\
\hline
Total number of time slots $T$ & $11$  \\
\hline
Total number of RBs $N_{\text{RB}, \text{total}}$  & $70$\\
\hline
Training batch size $B$  & $128$ \\
\hline
Initial learning rate $\eta$ of Adam Optimizer & $10^{-3}$ \\
\hline
Time window size $T_{c}$ & $3\,$ \\
\hline
\end{tabular}
\label{tab:tab002}
\end{table}

In our simulations, the BS serves $K = 10$ single-antenna UEs, which are randomly distributed within the $20 \times 20\,m^2$ service area.
Note that $v$ and $h$ are the speed and height of UEs, which are also randomly chosen from the ranges $[h_{\text{min}}, h_{\text{max}}]$ and $[0 , v_{\text{max}}]$, respectively.
Every UE moves straight\footnote{While UE mobility patterns can be complex in real-world scenarios, the proposed LMM-PS can predict UE trajectories every 100\,ms so that LMM-PS can maintain robust prediction performance even under random-walk-like mobility scenarios.} and changes the 
direction randomly if it reaches the edge of service area.
To generate the path gain $\alpha_{k, p} = \sqrt{\beta^{\text{L}}_{k, p}} \beta^{\text{S}}_{k, p} = \sqrt{\rho(r_{k, p}) \cdot 10^{\frac{z_{k} \sigma_{\text{sh}}}{10}}} \beta^{\text{S}}_{k, p}$ for the $p$-th path, we use the path loss model and shadow fading specified in 3GPP TR 38.901 Rel. 17~\cite{pathlossmodel}, where
$\rho(r) = - (31.84 + 21.5\,\text{log}_{10}(r) + 19\, \text{log}_{10}(f_c))$ is the path loss, $\sigma_{\text{sh}} = 4$, and $z_k \sim \mathcal{N}(0,1)$.
Note that $\beta^{\text{S}}_{k, p} \sim \mathcal{C}\mathcal{N}(0, 1)$ is the small-scale fading coefficient.
The system parameters are summarized in Table~\ref{tab:tab002}.

\begin{figure}[t]
	\centering
	\includegraphics[width=1.0\columnwidth, height = 7.0cm]{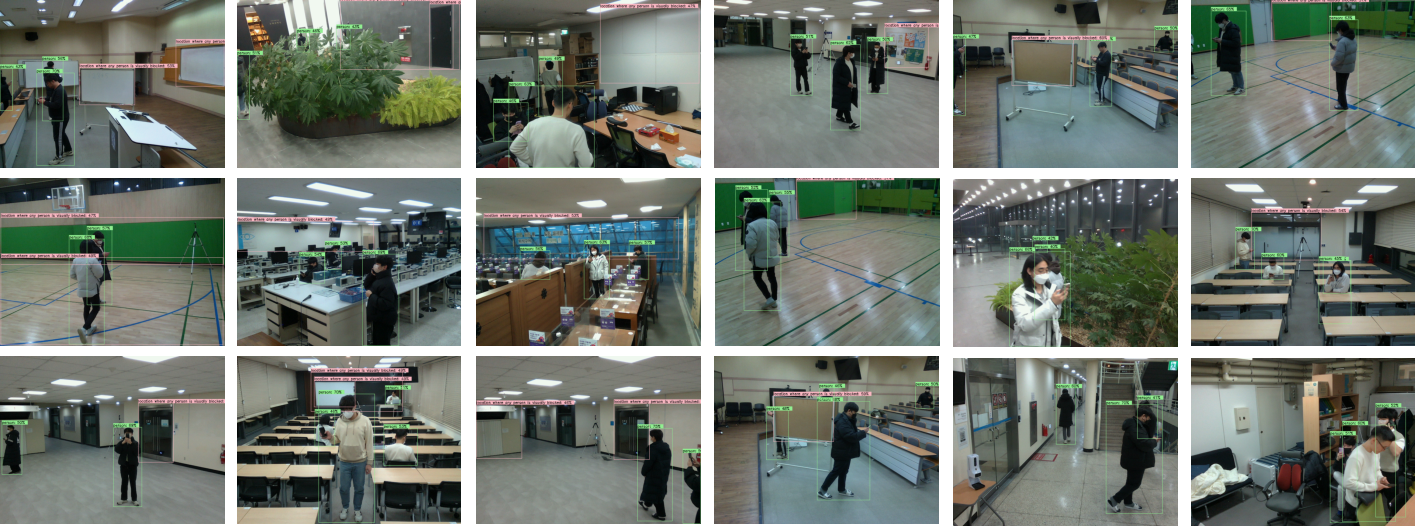}
	\vspace{-0.3cm}
	\caption{Object detection results of LMM-PS on the VOBEM2 set. Green and pink bounding boxes indicate persons and obstacles, respectively.}
 \vspace{-0.5cm}
 \label{fig:valresult}
\end{figure}
As for the pre-trained LLaVA model, we select LLaVA-v1.6-13B from the latest set of LLaVA models, which range from v1.6-7B to v1.6-34B, to balance inference speed and performance.
In the training phase of TCP-L and
 TCP-N, we use the Adam optimizer, a widely used optimizer that maintains the robustness of the learning process~\cite{adam}. 
Additionally, to help the smooth convergence of the loss value during training, we employ a learning rate decay strategy, reducing the learning rate by a factor of $10$ after every $100$ training epochs.


\begin{table*}[t]
    \centering
    \caption{Object detection and localization performance using GLIPv2 and the VOBEM2 dataset.}
    \vspace{-0.2cm}
    \resizebox{2.0\columnwidth}{!}{
    \begin{tabular}{c|c|cc|c|cc}
        \hline
        &
        \multicolumn{1}{c}{Human/Phone} & \multicolumn{2}{|c}{Localization error (Phone)} & \multicolumn{1}{|c}{Obstacle} & \multicolumn{2}{|c}{Localization error } \\
          & Recall (\%) & Distance (cm) & Az/El Angles (deg) & Precision/Recall (\%) & Distance (cm) & Az/El Angles (deg)  \\
         \hline
        GLIPv2 & \textbf{99.18\,/\,95.45}    & \textbf{0.84}    & \textbf{0.113\,/\,0.256}       & \textbf{90.00\,/\,89.55}   & \textbf{25.9}    & \textbf{0.366\,/\,0.687}      \\
         Swin Transformer & 98.09\,/\,92.56 & 1.36 & 0.133\,/\,0.289 & - & - & -\\
        EfficientDet-D8 & 94.55\,/\,90.08 & 3.20 & 0.140\,/\,0.400 & - & -  & - \\
         \hline
        \end{tabular}}
    \label{tab:vision_result}
\end{table*}
\begin{table*}[t]
    \centering
    \caption{Mobility and 
 blockage prediction performance using LLaVA-v1.6-13B. Latency and power consumption are estimated based on the Qualcomm AI 100 Edge.}
    \vspace{-0.2cm}
      \resizebox{2.0\columnwidth}{!}{
    \begin{tabular}{c|ccc|c|cc}
        \hline
         & \multicolumn{3}{|c}{Localization error (Phone)} & \multicolumn{1}{|c}{Blockage prediction}  
         & \multicolumn{2}{|c}{Resource usage}  
          \\
          & Distance (cm) & Az Angle (deg) & El Angle (deg)  & Accuracy (\%) & Latency (ms)  & Power (W) 
          \\
          \hline
        LLaVA-v1.6-13B & 12.78  & 0.27  & 0.1 & 91.7 & 25 & 15\\
        DL-based trajectory prediction~[B3] & 45.66  & 8.5  & 7.86 &- & 31 & 25 \\
        Sage-Husa filter-based trajectory prediction~[B4] & 22.01 & 14.26  & 13.66 & - & 30 & 20 \\
        DL-based blockage prediction & -  & -  & - & 82.4 & 31 & 25 \\
         \hline
        \end{tabular}
        }
          \vspace{-0.6cm}
    \label{tab:mobility_result}
\end{table*}
To evaluate the UE localization performance of GLIPv2, we use VOBEM2~\cite{VOBEM2}, 
a sensing dataset tailored for wireless communications.
VOBEM2 has 104 pairs
 of RGB and depth images containing persons and cell phones, some of which are obscured by
 obstacles.
In our
experiments, we define obstacles as walls, pillars, or boards.
For the obstacle detection evaluation, we manually annotate
bounding boxes for each obstacle in the VOBEM2 images (see https://github.com/islab-github/VOBEM2).
We then perform the object detection on these newly annotated images using GLIPv2 (see Fig.~\ref{fig:valresult}).
Furthermore, to evaluate the UE mobility and blockage prediction performances of LLaVA, we use synthetically generated test samples of TCP.

For the object detection performance comparison, we employ two off-the-shelf object detectors: 1) Swin Transformer~\cite{swin} and 2) EfficientDet-D8~\cite{Efficientdet}.
Also, for the mobility (trajectory) prediction performance comparison, we use two benchmark schemes:
\begin{enumerate}
\item DL-based trajectory prediction~\cite{DLtrajectoryprediction}: In this scheme, the BS uses an encoder-decoder architecture, where the encoder captures UEs' common movement patterns in a specific area from past  trajectories, and then the decoder predicts future trajectories based on these patterns.
\item Sage-Husa Filter-based trajectory prediction~\cite{Kalmantrajectoryprediction}: In this scheme, the BS estimates the unknown and time-varying noises in the measurements of UEs' movements based on the noise statistics. After that, the BS predicts the future trajectories using these noise estimates.
\end{enumerate}
To compare the scheduling performance, we use four benchmark schemes:
\begin{enumerate} 
\item  DRL-based preemptive scheduling~\cite{conv3_DL}: The BS exploits DRL to determine each UE's optimal MCS index and RB allocation for the $T$-th (future) slot. These scheduling decisions are made based on past estimated channels and SNRs of UEs.
\item Round Robin-based reactive scheduling~\cite{roundrobinscheduling}:  The BS allocates an equal number of RBs to each UE, which implies that rate requirements of some UEs may not be satisfied. Note that MCS indices are selected based on the SNRs in the $(T-1)$-th slot.
\item Maximum SNR-based reactive scheduling~\cite{maximumSNRscheduling}: The BS prioritizes RB allocation to UEs with high SNR values in the $(T-1)$-th slot. This process continues until all RBs are assigned, without considering fairness or QoS guarantees for all UEs.
\item Proportional fair-based reactive scheduling~\cite{proportionalfairscheduling}: The BS makes scheduling decisions based on UEs' SNRs observed in the $(T-1)$-th slot, allocating RBs to pursue a balance between the total system throughput and QoS requirement satisfaction. 
\end{enumerate}

\subsection{Simulation Results}\label{sec_5_2}


In Table~\ref{tab:vision_result}, we present the UE and cell phone detection performance in terms of recall (i.e., percentage of the detected objects among all target objects).
We observe that GLIPv2 outperforms Swin Transformer and EfficientDet, achieving an average recall of over 97\% for detecting both persons and cell phones.
By leveraging both image and text inputs, GLIPv2 understands an input sentence and identifies the image locations of the objects mentioned in the sentence, thereby enhancing the performance of conventional object detectors that rely solely on image inputs.

We also present the obstacle detection performance in Table~\ref{tab:vision_result}. 
The results show that GLIPv2 achieves around 90\% in both precision (i.e., the percentage of the correctly detected objects among total detected objects) and recall in the obstacle detection. 
Since conventional object detectors are trained to identify pre-defined object classes only, they struggle to detect obstacles that were not explicitly labeled or included in their training data.
 





In Table IV, 
 we evaluate the mobility prediction performance of  LLaVA-v1.6-13B.
We observe that LLaVA accurately predicts large-scale LoS channel parameters (distance and angles), achieving an average distance error of less than 15\,cm. 
Furthermore, we compare the blockage prediction accuracy of the proposed LMM-PS with a DL-based blockage prediction scheme~\cite{blockageprediction_conv}, which uses a long short-term memory (LSTM) network to predict future blockage status based on a sequence of past estimated channels.
The results in Table IV show that LMM-PS, which exploits LLaVA's mobility prediction results, improves blockage prediction accuracy by over 10\% compared to the LSTM-based scheme, thanks to its ability to quickly identify the surrounding wireless environment through the visual sensing information.

\begin{figure*}[t]
	\centering
    \subfloat[]{\includegraphics[width=1.0\columnwidth, height=7.5cm]{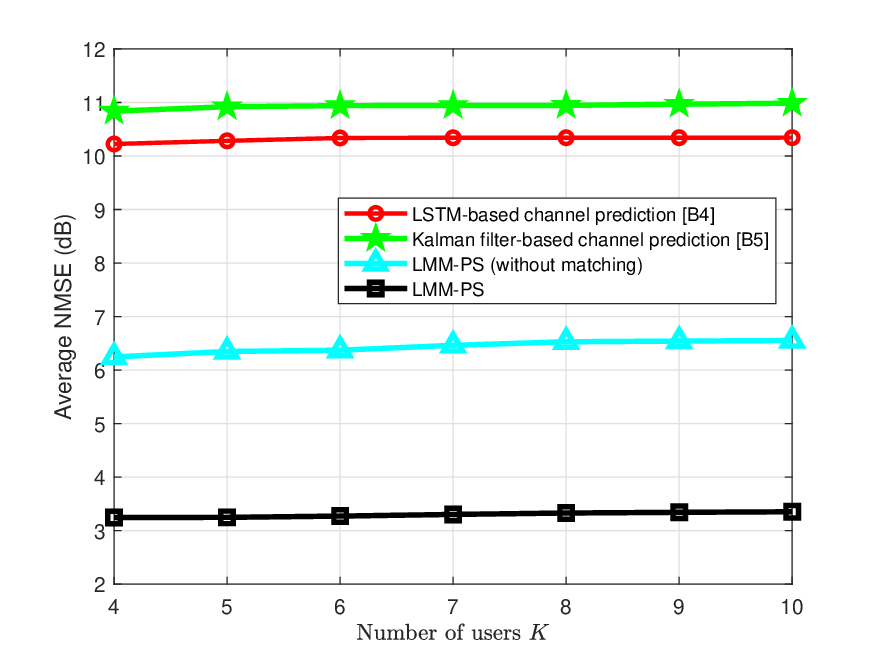}}
     \subfloat[]{\includegraphics[width=1.0\columnwidth, height=7.5cm]{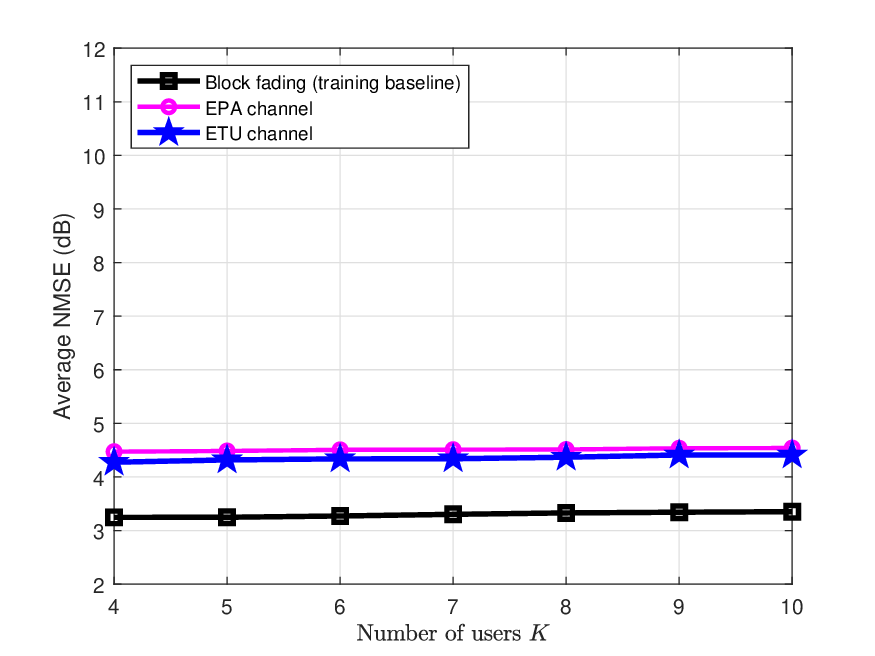}}
	\caption{(a) Average NMSE of channel prediction techniques and (b) Average NMSE of LMM-PS for various channel models.}
	\label{fig:channelpredictionresults}
\end{figure*}

 We also see from Table IV that the latency and power consumption of LMM-PS are much lower than those of the conventional techniques.
Since future UE positions are predicted from captured images using LLaVA, in essence the beam sweeping latency of conventional RF-based trajectory prediction techniques is replaced with the LMM inference latency, leading to a reduction of the latency and power consumption of LMM-PS.
For example, when employing the latest AI-focused SoC \textit{Qualcomm AI 100 Edge}, inference time is around 25\,ms, but the beam sweeping latency of conventional RF-based  techniques is about 30\,ms~\cite{YJAhnmag}. 
Similarly, in terms of the power consumption, the beam sweeping stage typically consumes 20\,W~\cite{YJAhnmag} while the \textit{Qualcomm AI 100 Edge} consumes around 15\,W, indicating promising energy savings with LMM-PS.


To validate the effectiveness of the proposed Transformer-aided models in LoS path gain and NLoS channel parameter estimation, we have newly investigated the channel prediction performance of LMM-PS in terms of the average NMSE (see Fig. 12(a)).
We have also conducted the performance evaluation of two recent approaches: \cite{DLchannelprediction}~LSTM-aided channel prediction and \cite{Kalmanchannelprediction}~Kalman filter-based channel prediction (see \cite{DLchannelprediction} and \cite{Kalmanchannelprediction} in Fig. 12(a)).
To ensure a fair comparison, we used the same large-scale LoS channel parameters (distance and angles of the
LoS path) predicted by the LMM as inputs for LoS channel prediction in all schemes we tested.
From Fig. 12(a),  we observe that LMM-PS performs uniformly better than~\cite{DLchannelprediction} and~\cite{Kalmanchannelprediction}.
Since Transformer assigns relatively large attention weights
to input data (i.e., past channels) that are highly correlated to the
output values (i.e., future channel parameters), LMM-PS can effectively capture long-term spatial correlation that tends to be preserved even under sudden environmental changes, which leads to a meaningful improvement in the channel prediction process.
 
 \begin{figure*}[t]
	\centering
    \subfloat[]{\includegraphics[width=1.0\columnwidth, height=7.5cm]{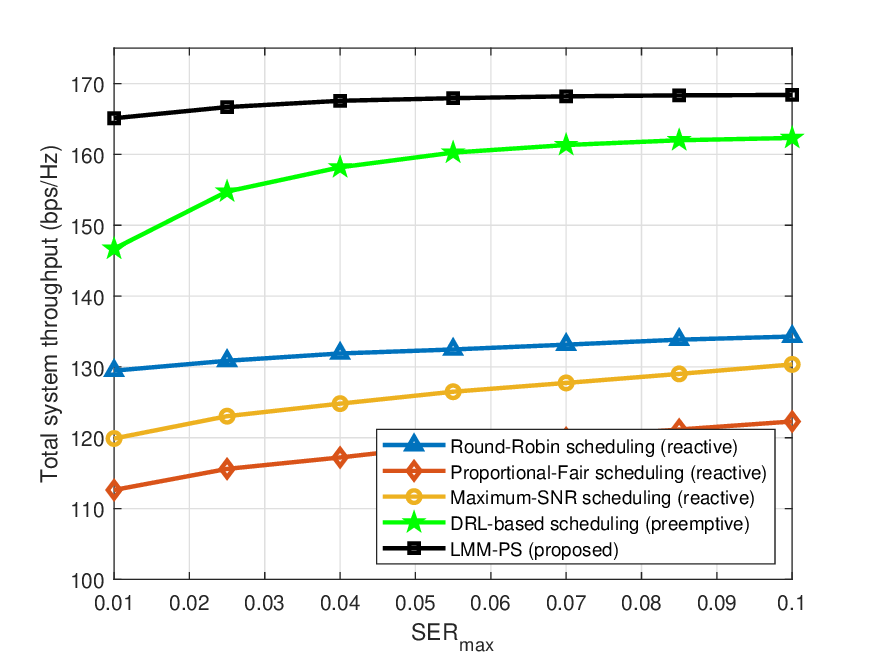}}
     \subfloat[]{\includegraphics[width=1.0\columnwidth, height=7.5cm]{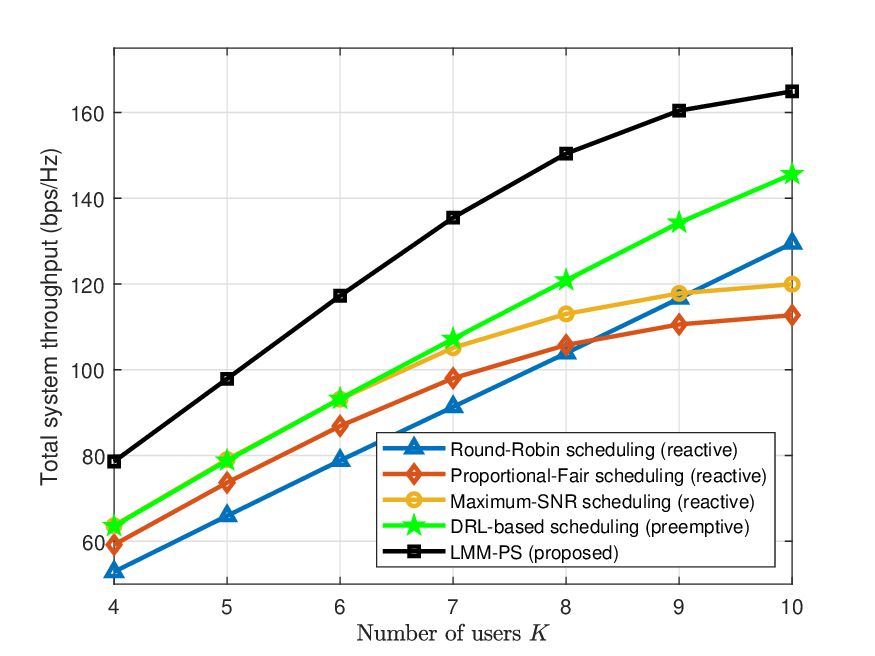}}
	\caption{(a) Total system throughput as a function of $\text{SER}_{\text{max}}$ ($K = 10$, $v = 12\text{km/h}$) and (b) Total system throughput as a function of number of UEs ($\text{SER}_{\text{max}} = 10^{-2}$, $v = 12\text{km/h}$).}
	\label{fig:originalResults}
\end{figure*}

To demonstrate the efficacy of LMM-PS under various channel conditions, we have tested its channel prediction performance for the extended pedestrian-A (EPA) and extended typical urban (ETU) channels (see Fig. 12(b)).
We observe that the overall performance of LMM-PS under these models is similar to that under the block-fading channel. 
For example, when
 $K = 4$, LMM-PS under ETU channel achieves an average NMSE of about 4.27\,dB, which is comparable to the NMSE achieved by LMM-PS under block-fading channel (3.25\,dB).

In Fig. 13(a), we plot the total system throughput of LMM-PS as a function of $\text{SER}_{\text{max}}$.
We confirm that LMM-PS outperforms the conventional techniques across the board.
For example, LMM-PS achieves 11\% throughput gain over the DRL-based scheduling at $\text{SER}_{\text{max}} = 10^{-2}$. 
This is mainly because LMM-PS predicts the future movements of UEs using LLaVA and the channel parameters for strong NLoS paths using Transformers, which helps LMM-PS predict future LoS and NLoS channel component used to compute the future PF metric.

In Fig. 13(b), we plot the total system throughput of LMM-PS as a function of the number of UEs $K$.
We observe that LMM-PS achieves significantly higher throughput compared to conventional techniques. While conventional techniques relying exclusively on past estimated channels often fail to capture sudden channel changes (e.g., blockages) in the wireless environment, LMM-PS can understand these events using the visual sensing information and LMMs, resulting in substantial total system throughput improvement.
For example, when $K = 7$, LMM-PS achieves a 32\% throughput gain over the Round-Robin scheduling technique.


To judge the generalization ability of LMM-PS, we newly evaluated UE and obstacle localization performance using the VOMM dataset [3], a vision dataset tailored for mobility scenarios (see Table V). 
For the obstacle detection evaluation, we manually annotated bounding boxes for each obstacle in the VOMM images (see
https://github.com/islab-github/VOMM).
From our experiments, we confirm that the localization performance of LMM-PS in VOMM is comparable to that in VOBEM2. Based on the localization results in Table V, we further evaluated the throughput performance of LMM-PS
under various mobility scenarios (see Fig. 14(a)). We observe that the throughput performance of LMM-PS in VOMM is comparable to that in VOBEM2. For example, when
the UE velocity is $v = 12\text{km/h}$, LMM-PS in VOMM achieves about 167.8 bps/Hz, which is very close to the total system throughput achieved in VOBEM2 (168.4 bps/Hz).

Finally, to assess the robustness of LMM-PS in various environments, we have designed new simulation spaces (using AutoCAD) by deploying obstacles with varying densities $\rho_o$,  which is defined as the ratio of the $XY$-plane area occupied by obstacles to the total service area.  
We then tested the throughput performance of LMM-PS in these spaces (see Fig. 14(b)).
We observe that LMM-PS achieves significant improvement in the total system throughput over the conventional strategies.
For instance, when $\rho_{o} = 20\%$, LMM-PS achieves 26\% gain in the total system throughput over the Round Robin-based technique.
\begin{table*}[t]
    \centering
      \caption{Object detection and localization performance using GLIPv2 and the VOMM dataset.}
    \vspace{-0.2cm}
    \resizebox{2.0\columnwidth}{!}{
    \begin{tabular}{c|c|cc|c|cc}
        \hline
        &
        \multicolumn{1}{c}{Phone} & \multicolumn{2}{|c}{Localization error (Phone)} & \multicolumn{1}{|c}{Obstacle} & \multicolumn{2}{|c}{Localization error } \\
          & Recall (\%) & Distance (cm) & Az/El Angles (deg) & Precision/Recall (\%) & Distance (cm) & Az/El Angles (deg)  \\
         \hline
        GLIPv2 & \textbf{99.57}    & \textbf{0.67}    & \textbf{0.090\,/\,0.223}       & \textbf{92.28\,/\,88.67}   & \textbf{46.8}    & \textbf{0.569\,/\,0.929}      \\
        Swin Transformer & 95.26 & 3.64 & 0.142\,/\,0.400 & - & -  & - \\
         \hline
        \end{tabular}}
    \label{tab:vomm_result}
    \vspace{-0.5cm}
\end{table*}
 \begin{figure*}[t]
	\centering
    \subfloat[]{\includegraphics[width=1.0\columnwidth, height=7cm]{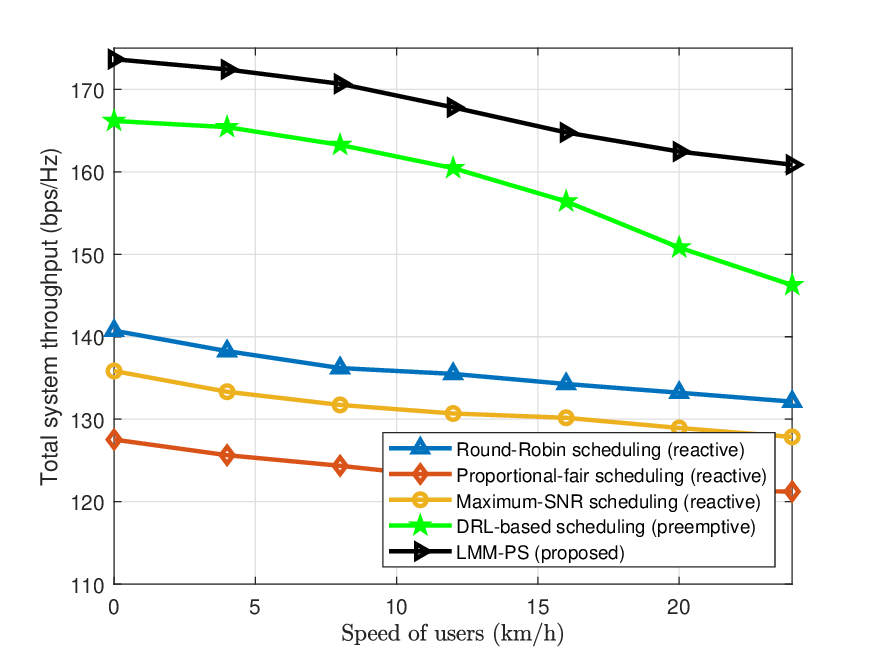}}
     \subfloat[]{\includegraphics[width=1.0\columnwidth, height=7cm]{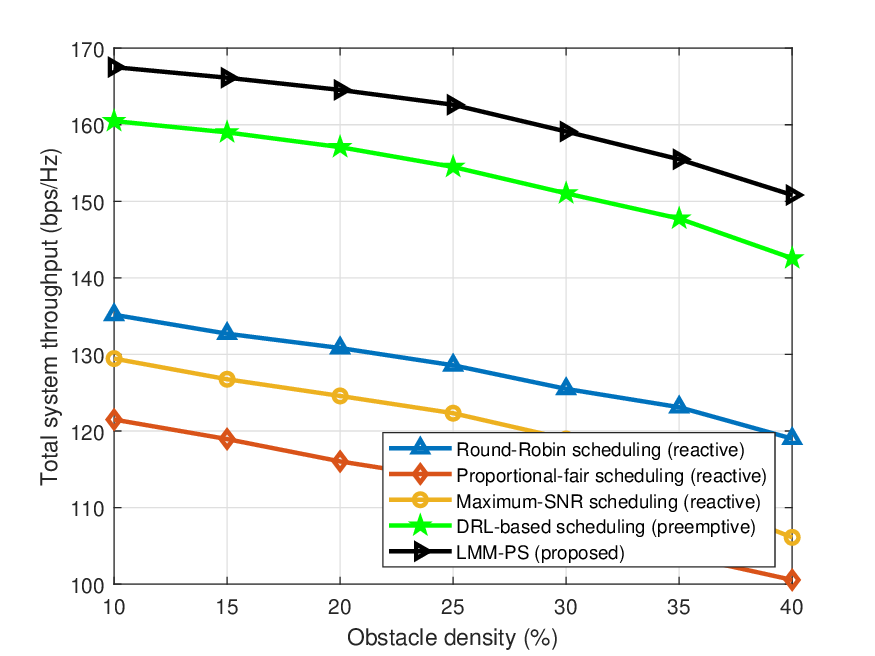}}
	\caption{(a) Total system throughput as a function of UE velocity ($K = 10$, $\text{SER}_{\text{max}} = 10^{-1}$) and (b) Total system throughput as a function of the obstacle density ($K = 10$, $\text{SER}_{\text{max}} = 10^{-1}$, $v = 12\text{km/h}$).}
	\label{fig:originalResults}
\end{figure*}

\section{Conclusion}
In this paper, we proposed an LMM-aided preemptive scheduling technique called LMM-PS for future 6G autonomous communication systems. 
Unlike conventional approaches relying solely on RF signals, LMM-PS exploits  visual sensing information and LMMs to quickly detect and analyze the signal propagation environment.
By forecasting the future positions of mobile devices and sudden events, and combining these with past RF-based channel information, LMM-PS can generate accurate future channel estimates for the preemptive scheduling.
Through numerical experiments, we demonstrated that LMM-PS significantly outperforms traditional RF-based scheduling techniques in terms of the total system throughput.
While we did not use latest LMMs (e.g., GPT-4V and LLaMA-VID) in the current version of LMM-PS, replacing the existing backbones with these models that handle new input types (e.g., image sequences) will be a promising research direction.
We believe that the proposed LMM-PS is an effective tool to support the emerging LLM/LMM-based autonomous communication paradigm, where tasks are executed through the cooperation of AI, multimodal sensing, and wireless communications.


\bibliographystyle{IEEEtran}

\begin{thebibliography}{1}

    \bibitem{conference}
    S. Kim, J. Son, and B. Shim, ``Large multimodal model-based scheduling for autonomous communication systems," to appear in {\em IEEE Veh. Technol. Conf. (VTC)}, 2025.
    
    \bibitem{generativeAI}
     Y. Chang \textit{et al.}, “A survey on evaluation of large language models,” {\em ACM Trans. Intell. Syst. Technol.}, vol. 15, no. 3, pp. 1–45, Mar. 2024.

     \bibitem{LMM}
     M. Xu \textit{et al.}, “A survey of resource-efficient llm and multimodal foundation models," {\em arXiv preprint arXiv:2401.08092}, 2024.

   

    \bibitem{autonomousdrivingLMM}
    Z. Yang, X. Jia, H. Li, and J. Yan, “LLM4Drive: A survey of large language models for autonomous driving," {\em arXiv preprint arXiv:2311.01043}, 2023.

     \bibitem{humanoidLMM}
     M. Ahn \textit{et al.}, “AutoRT: Embodied foundation models for large scale orcehstration of robotic agents," {\em arXiv preprint arXiv:2401.12963}, 2024.
     
    \bibitem{surgeryLMM}
    G. Wang  \textit{et al.}, “Surgical-LVLM: Learning to adapt large vision-language model for grounded visual question answering in robotic surgery," {\em arXiv preprint arXiv:2401.08092}, 2024.
    
    \bibitem{motivation1} 
     H. Zhou \textit{et al.}, “Large language model (LLM) for telecommunications: A
    comprehensive survey on principles, key techniques, and opportunities,”
    \textit{IEEE Commun. Surveys Tuts.}, vol. 27, no. 3, pp. 1955–2005, Jun. 2025.
    
    \bibitem{motivation2} F. Jiang, Y. Peng, L. Dong, K. Wang, K. Yang, C. Pan, D. Niyato, and O. A. Dobre,
    “Large language model enhanced multi-agent systems for 6G communications,” \textit{IEEE Wireless Commun.}, vol. 31, no. 6, pp. 48-55, Dec. 2024.

       
    \bibitem{YJAhn_SNRdrop}  
    Y. Ahn, J. Kim, S. Kim, S. Kim, and B. Shim, "Sensing and computer vision-aided mobility management for 6G millimeter and terahertz communication systems," {\em IEEE Trans. Commun.}, vol. 72, no. 10, pp. 6044-6058, Oct. 2024.
    


    \bibitem{conv1_preemptive}
    R. Zeng, T. Liu, X. Yu, and Z. Zhang, "Novel channel quality indicator prediction scheme for adaptive modulation and coding in high mobility environments,"
     {\em IEEE Access}, vol. 7, pp. 11543-11553, 2019.

    \bibitem{conv2_preemptive}
    A. Balieiro, K. Dias and P. Guarda, "Addressing the CQI feedback delay in 5G/6G networks via machine learning and evolutionary computing," {\em Intelligent and Converged Networks}, vol. 3, no. 3, pp. 271-281, Sept. 2022.

    \bibitem{conv3_preemptive}
    Q. He, G. Dan, and G. P. Koudouridis, ‘‘Semi-persistent scheduling for 5G downlink based on short-term traffic prediction,’’ in {\em Proc. IEEE Global Commun. Conf. (GLOBECOM)}, 2020.

    \bibitem{transformers}
    A. Vaswani \textit{et al.}, ``Attention is all you need,"
    {\em Adv. Neural Inform. Process. Syst.}, 2017. 
    
   
    \bibitem{MCSconstraint}
    3GPP Technical Specification 38.212, 
     “5G; NR; Multiplexing and channel coding,”
    {\em v15.7.0}, Sept. 2019. 
    
     \bibitem{5GNRsynchronization}
    A. Omri, M. Shaqfeh, A. Ali, and H. Alnuweiri, “Synchronization procedure in 5G NR systems,” {\em IEEE Access}, vol. 7, pp. 41286--41295, 2019.

    \bibitem{snkim_channelmodel}
    S. Kim, J. Wu, and B. Shim, “Efficient channel probing and phase shift control for mmWave reconfigurable intelligent surface-aided communications,” {\em  IEEE. Trans. Wireless Commun.}, vol. 23, no. 1, pp. 231--246, Jan. 2024.

    \bibitem{jwson_channelcomponent}
   J. Son, I. Keum, H. Kim, H. Cho, and B. Shim,
   “Transformer-based environment-aware localization in the NLoS scenarios," in {\em Proc. IEEE Wireless Commun. Netw. Conf. (WCNC)}, 2024.

     \bibitem{conv4_DL}
     X. Ye, Y. Yu, and L. Fu, “Deep reinforcement learning based link adaptation technique for LTE/NR Systems," {\em IEEE Trans. Veh. Technol.}, vol. 72, no. 6, pp. 7364--7379, Jun. 2023. 

   \bibitem{sJeong}
    S. Jeong, H. Ju, S. Kim, and B. Shim, ``Automated environment-aware channel feedback for 6G massive MIMO systems," 
    in {\em Proc. IEEE Glob. Commun. Wkshps. (GC Wkshps)}, 2023.
 
     
    \bibitem{conv3_DL}
     Y. Liao, Z. Yang, Z. Yin, and X. Shen, “DQN-based adaptive MCS and SDM for 5G massive MIMO-OFDM downlink," {\em IEEE Commun. Lett.}, vol. 27, no. 1, pp. 185--189, Jan. 2023. 

       
    \bibitem{PMItoCQItableandMCSordertable}
    3GPP Technical Specification 38.214, 
     “5G; NR; Physical layer procedures for control,”
    {\em v15.6.0}, Jun. 2019. 
    
    \bibitem{CQItoMCStable}
    3GPP Technical Specification 38.214, 
     “5G; NR; Physical layer procedures for data,”
    {\em v16.2.0}, Jul. 2020. 

  
     \bibitem{conv2_DL}
     E. Bobrov, D. Kropotov, H. Lu, and D. Zaev, “Massive MIMO adaptive modulation and coding using online deep learning algorithm,” {\em IEEE Commun. Lett.}, vol. 26, no. 4, pp. 818--822, Apr. 2022.

    \bibitem{blockageandthreshold}
    Y. Ahn, J. Kim, S. Kim, and B. Shim, ``Computer vision-aided proactive mobility management for 6G terahertz communications," in {\em Proc. IEEE Glob. Commun. Conf. (GLOBECOM)}, 2023.
    
    \bibitem{CQIfeedbackdelayandperiod}
    3GPP Technical Specification 38.211, 
     “5G; NR; Physical layer procedures for data,”
    {\em v16.2.0}, Jun. 2018. 

     \bibitem{maximumSNRscheduling}
    H. Oh and H. Nam, “Maximum rate scheduling with adaptive modulation in mixed impulsive noise and additive white gaussian noise environments,” {\em IEEE Trans. Wireless Commun.}, vol. 20, no. 5, pp. 3308–3320, May 2021.

   
    \bibitem{SNRtoSER}
    A. J. Goldsmith, {\em Wireless Communication.} Cambridge, U.K: Cambridge Univ. Press, 2005.
    
     \bibitem{SDM}
     Y. Chen, Y. Wu, Y. T. Hou, and W. Lou, “mCore: Achieving submillisecond scheduling for 5G MU-MIMO systems,” in {\em Proc. IEEE
     Conf. Comput. Commun.}, 2021.
     
    \bibitem{glipv2}
    H. Zhang, P. Zhang, X. Hu, Y-C. Chen, L. H. Li, X. Dai, L. Wang, L. Yuan, J-N. Hwang, and J. Gao, “GLIPv2: Unifying localization and vl understanding,” in {\em Proc. Adv. Neural Inf. Process. Syst.}, vol. 35, 2022.

    
    \bibitem{LLaVA}
    H. Liu, C. Li, Y. Li, Y. Lee, “Improved baselines with visual instruction tuning," in {\em Proc. Comput. Vis. Pattern Recognit. (CVPR)}, pp. 26296--26306, 2024.
    
    \bibitem{lora}
    E. J. Hu, Y. Shen, P. Wallis, Z. Allen-Zhu, Y. Li, S. Wang, L. Wang, and W. Chen,  “Lora: Low-rank adaptation of large language models,” in \textit{Proc. Int. Conf. Learn. Represent. (ICLR)}, 2021.
    
     \bibitem{PFmetric}
    Y. Barayan and I. Kostanic, “Performance evaluation of proportional fairness scheduling in LTE,” in {\em Proc. World Congr. Eng. Comput. Sci.}, vol. 2,  pp. 712--717, 2013.
    
    \bibitem{vit}
    D. Alexey \textit{et al.},  "An image is worth 16x16 words: Transformers for image recognition at scale," in 
    {\em  arXiv preprint arXiv:2010.11929}, 2020.

    \bibitem{bert}
    J. Devlin, M. Chang, K. Lee, and K. Toutanova,
    “Bert: Pre-training of deep bidirectional transformers for language understanding,” in 
    {\em arXiv preprint arXiv:1810.04805}, 2019.

    \bibitem{structure_of_LMMs}
    S. Song, X. Li, S. Li, S. Zhao, J. Yu, J. Ma, X. Mao, W. Zhang,  and M. Wang, “How to bridge the gap between modalities: Survey on multimodal large language model,” {\em IEEE Trans. Knowl. Data Eng.},  vol. 37, no. 9, pp. 5311--5329, Sept. 2025.


     \bibitem{vomtc}
    S. Kim, Y. Ahn, D. Park, and B. Shim, ``VOMTC: Vision objects for millimeter and terahertz communications," {\em IEEE Trans. Cogn. Commun. Netw.}, vol. 11, no. 1,  pp. 243--257, Feb. 2025.

        
    \bibitem{beamforming}
    S. Kim, J. Kim, J. Lee, and B. Shim, 
    “Computer vision-aided beamforming for millimeter and terahertz communications,” \textit{Jour. Kor. Instit. Commun. Inf. Sci.}, vol. 49, no. 12, pp. 1685--1694, Dec. 2024.

     \bibitem{Hungarian}
    H. W. Kuhn,  “The hungarian method for the assignment problem,” {\em Nav. Res. Logistics (NRL)}, vol. 52, no. 1, pp. 
    7--21, Feb. 2005.

    \bibitem{codebook_based_paramacquisition}
    Y. Liu, J. Wu, S. Kim, and B. Shim,  “Vision-aided blockage prediction and proactive handover for indoor mmwave and terahertz communications,” in {\em Proc. IEEE Glob. Commun. Conf. (GLOBECOM)}, 2023.
    
    \bibitem{pathlossmodel}
    3GPP TR 38.901, “Study on channel model for frequencies from 0.5 to 100 GHz,"  {\em v17.0.0}, 2022.
   
        
   
    \bibitem{adam}
    D. P. Kingma and J. Ba, “Adam: A method for stochastic optimization,” {\em arXiv preprint arXiv:1412.6980}, 2014.
    
    
     \bibitem{VOBEM2}
    S. Kim, J. Moon, J. Kim, Y. Ahn, D. Kim, S. Kim, K. Shim, and B. Shim, ``Role of sensing and computer vision in 6G wireless communications," 
    \textit{IEEE Wireless Commun.},
     vol. 31, no. 5, pp. 264--271, Oct. 2024.


    \bibitem{swin}
    Z. Liu, Y. Lin, Y. Cao, H. Hu, Y. Wei, Z. Zhang, S. Lin, and B. Guo, ``Swin Transformer: Hierarchical vision transformer using shifted windows," in {\em Proc. IEEE/CVF Int. Conf. Comput. Vis. (ICCV)}, pp. 10012--10022, 2021.

    \bibitem{Efficientdet}
    M. Tan, R. Pang, and Q. V. Le, ``Efficientdet: Scalable and efficient object detection," in {\em Proc. Conf. Comput. Vis. Pattern Recognit. (CVPR)}, pp. 10781-10790, 2020.       
    \bibitem{DLtrajectoryprediction}
    \noindent C. Wang, L. Ma, R. Li, T. S. Durrani, and H. Zhang, ``Exploring trajectory prediction through
    machine learning methods," \textit{IEEE Access}, vol. 7, pp. 101441-101452, 2019.
    \bibitem{Kalmantrajectoryprediction} Y. Yang and W. Gao, ``An optimal adaptive Kalman filter,'' \textit{J. Geodesy}, 2006.
    \bibitem{roundrobinscheduling}
    H. M. Chaskar and U. Madhow, ‘‘Fair scheduling with tunable latency: A round-robin approach,’’ {\em IEEE/ACM Trans. Netw.}, vol. 11, no. 4, pp. 592–601, Aug. 2003.

   
    \bibitem{proportionalfairscheduling}
    R. Kwan, C. Leung, and J. Zhang, “Proportional fair
    multiuser scheduling in LTE,” {\em IEEE Signal Process.
    Lett.}, vol. 16, no. 6, pp. 461–464, Jun. 2009.  
     \bibitem{blockageprediction_conv}
    M. Alrabeiah and A. Alkhateeb, “Deep learning for mmWave beam and blockage prediction using sub-6 GHz channels,” {\em IEEE Trans. Commun.}, vol. 68, no. 9, pp. 5504–5518, Sept. 2020.
   
    \bibitem{YJAhnmag}
    Y. Ahn, J. Kim, S. Kim, K. Shim, J. Kim, S. Kim, and B. Shim, "Towards intelligent millimeter and terahertz communication for 6G: Computer vision-aided beamforming, \textit{IEEE Wireless Commun.}, vol. 30, no. 5, pp. 179-186, Oct. 2023.    
   \bibitem{DLchannelprediction} Y. Yang, D. B. Smith, and S. Seneviratne, “Deep learning channel prediction for transmit power control in wireless body
    area networks,” in \textit{Proc. IEEE Int. Conf. Commun.}, 2019.
    \bibitem{Kalmanchannelprediction}  H. Kim, S. Kim, H. Lee, C. Jang, Y. Choi, and J. Choi, ``Massive MIMO channel prediction: Kalman filtering vs. machine
    learning," \textit{IEEE Trans. Commun.}, vol. 69, no. 1, pp. 518--528, Jan. 2021.
    
 
\end{thebibliography}

\end{document}